%% file: LatticeHiggs1.tex
\documentclass[12pt]{article}
\usepackage{sint_modi,cite}

\usepackage{amssymb}
\usepackage{epsfig}

\def\a{\alpha}
\def\b{\beta}
\def\g{\gamma}
\def\G{\Gamma}
\def\D{\Delta}
\def\d{\delta}
\def\e{\varepsilon}

\def\s{\sigma}

\def\beq{\begin{equation}}
\def\eeq{\end{equation}}
\def\beqn{\begin{eqnarray}}
\def\eeqn{\end{eqnarray}}
\def\ba{\begin{eqnarray}}
\def\ea{\end{eqnarray}}

\def\m{{\tt -}}

\def\xprim2bar{\overline{x}^{\prime\prime}}

\def\beq{\begin{equation}}
\def\eeq{\end{equation}}
\def\tr{{\bf tr}}

\newcommand{\beqa}{\begin{eqnarray}}
\newcommand{\eeqa}{\end{eqnarray}}

\let\a=\alpha   \let\b=\beta   \let\g=\gamma   \let\d=\delta
\let\e=\epsilon         
        \let\m=\mu
\let\n=\nu                 
\let\s=\sigma        

\let\G=\Gamma  \let\D=\Delta

\let\a=\alpha   \let\b=\beta   \let\g=\gamma   \let\d=\delta
\let\e=\epsilon         
        \let\m=\mu
\let\n=\nu                 
\let\s=\sigma        

\let\G=\Gamma  \let\D=\Delta

\newcommand{\be}{\begin{equation}}
\newcommand{\ee}{\end{equation}}
\newcommand{\bea}{\begin{eqnarray}}
\newcommand{\eea}{\end{eqnarray}}

\usepackage{graphics}
\usepackage{graphicx}
\def\tr{{\rm tr}}
\newcommand{\eq}[1]{Eq.~(\ref{#1})}
\newcommand{\fig}[1]{Fig.~\ref{#1}}
\newcommand{\tab}[1]{Table~\ref{#1}}
\newcommand{\sect}[1]{Section~\ref{#1}}

\def\mO{\mathcal{O}}

\def\A5{(A_5)_{\rm lat}}
\def\thintablerule{\hrule height0.4pt}
\begin{document}
%%%%%%%%%%%%%%%%%%%%%%%%%%%%

%\vbox{\vskip0.0cm}
\rightline{CERN-PH-TH/2006-192}

\vskip 1.5cm
\centerline{\LARGE Lattice Gauge Theory Approach to Spontaneous}
\vskip 0.3 true cm
\centerline{\LARGE Symmetry Breaking from an Extra Dimension}
\vskip 0.6 true cm
\centerline{\large Nikos Irges}
\vskip1ex
\centerline{\it Department of Physics and Institute of Plasma Physics,}
\centerline{\it University of Crete, GR-710 03 Heraklion, Crete, Greece}
\centerline{\it e-mail: {\tt irges@physics.uoc.gr}}
\vskip 0.3 true cm
\centerline{\large Francesco Knechtli}
\vskip1ex
\centerline{\it CERN, Physics Department, TH Division, 1211 Geneva 23, Switzerland}
\centerline{\it e-mail: {\tt knechtli@mail.cern.ch}}
\vskip 0.8 true cm
\thintablerule
\vskip 2.0ex
\leftline{\bf Abstract}
We present lattice simulation results corresponding to an $SU(2)$ pure gauge theory
defined on the orbifold space $E_4 \times I_1$, where $E_4$ is the 
four-dimensional Euclidean space and $I_1$ is an interval,
with the gauge symmetry broken to a $U(1)$ subgroup at the two ends of the
interval by appropriate boundary conditions. 
We demonstrate that the $U(1)$ gauge boson acquires a mass from a Higgs mechanism. 
The mechanism is driven by two of the
extra-dimensional components of the five-dimensional gauge field which play 
respectively the role of 
the longitudinal component of the gauge boson and a massive real physical scalar,
the Higgs particle. Despite the non-renormalizable nature of the theory,
we observe only a mild cut-off dependence of the physical observables.
We also show evidence that there is a region in the parameter space where the
system behaves in a way consistent with dimensional reduction.

\vskip 1.0ex\noindent
\vskip 2.0ex
\thintablerule

\vskip-0.2cm

%%%%%%%%%%%%%%%%%%%%%%%%%%%%
\input{sect1.tex}

\input{sect2.tex}
\input{sect3.tex}
\input{sect4.tex}

\input{sect5.tex}
%%%%%%%%%%%%%%%%%%%%%%%%%%%%

\bigskip

{\bf Acknowledgement.}
We thank B. Bunk for his help in the construction of the programming code.
We are grateful to P. de Forcrand, M. Della Morte, P. Hasenfratz, Y. Hosotani,
M. L\"uscher and U.-J. Wiese
for discussions and helpful suggestions.
We thank the Swiss National Supercomputing Centre (CSCS) in Manno
(Switzerland)
for allocating computer resources to this project.
N. Irges thanks CERN for hospitality.

\bigskip

%%%%%%%%%%%%%%%%%%%%%%%%%%%%
\begin{appendix}
\input{appa.tex}
\input{appb.tex}
\end{appendix}
%%%%%%%%%%%%%%%%%%%%%%%%%%%%

\bibliography{latthiggs}           %or whatever your .bib file is
\bibliographystyle{h-elsevier.bst}   %if you use h-elsevier.bst

%%%%%%%%%%%%%%%%%%%%%%%%%%%%
\end{document}

%% file: sect1.tex
%%%%%%%%%%%%%%%%%%%%%%%%%%%%%%%%%%
\section{Introduction \label{s_intro}}
%%%%%%%%%%%%%%%%%%%%%%%%%%%%%%%%%%

Spontaneous symmetry breaking (SSB) is the phenomenon where the ground state of a system
does not access all of its available symmetry, apparently breaking the symmetry
group to a subgroup. In the Standard Model (SM) this is a crucial mechanism and it is not only
responsible for predicting the existence of a fundamental scalar field, the Higgs particle, but also 
for the gauge bosons and fermions acquiring a mass. The somewhat unsatisfactory fact about this
mechanism in the SM, is that the Higgs potential, which is the concrete object that drives SSB, is input by
hand at tree level in the Lagrangian, simply because we do not have any more fundamental way to generate it. 
There are many ideas of course trying to suggest an origin
for the Higgs and its potential, one of the most elegant being that the Higgs field is the extra dimensional
component of a higher dimensional gauge field
and that the potential is generated quantum mechanically \cite{Coleman:1973jx}.
The earliest scenarios considered as extra-dimensional space the sphere $S^2$
\cite{Fairlie:1979at,Fairlie:1979zy,Manton:1979kb,Forgacs:1979zs}.
In later applications
the extra-dimensional space was taken to be non-simply connected, like $S^1$ or
$T^2$ \cite{Hosotani:1983xw,Hosotani:1989bm,Antoniadis:2001cv}, so that 
the (non-contractible) Polyakov loops are non-trivial.
This is the general context where we would like to put ourselves in the present work.
Alternative ways to achieve SSB in gauge theories with extra dimensions are
discussed in \cite{Dvali:2001qr,Arkani-Hamed:2001nc}.

There are enough motivations to take this idea seriously besides
the economic way of generating the Higgs with its potential
but the property that drew a lot of recent attention to these theories is 
the so claimed attractive possibility of all order finiteness of the physical scalar mass.
This sounds like a paradox from the beginning since the very
point which has kept many field theorists rather hesitant from taking
such an idea seriously is that higher (than four) dimensional gauge theories
are non-renormalizable and in a typical non-renormalizable theory one would expect
that a mass parameter receives quantum corrections appearing in an arbitrary
power of some dimensionless quantity built out of a dimensionful coupling and the cut-off. 
This is to be compared with the renormalizable SM where the 
couplings are dimensionless and the Higgs mass receives only a quadratic
ultra-violet (UV) cut-off dependence under quantum corrections. 
Since even this quadratic UV sensitivity has been viewed as a drawback, 
supersymmetric generalizations of the SM were introduced and analyzed in
detail, where the power like cut-off sensitivity is not present due to cancellations
of infinities between superpartners. 
There is no doubt that supersymmetry is an elegant solution to the problem but it could 
happen that it is not realized at energies accessible in near future collider
experiments so it is useful to be aware of alternative solutions.   
Back then to extra dimensions, in the case where the extra (fifth here) dimension is 
compactified on a circle, one can carry out a one-loop calculation of the Higgs mass 
and verify its aforementioned finiteness 
\cite{vonGersdorff:2002as,Cheng:2002iz} and can 
even give all order arguments to that effect
\cite{Arkani-Hamed:2001mi,Masiero:2001im,vonGersdorff:2002rg,vonGersdorff:2002us,Irges:2004gy,Martinelli:2005ix,Lim:2006bx,Hosotani:2006nq}, 
but the problem with this solution is that
a simple circle compactification can not be realistic for various reasons, the absence of
chiral fermions being one of the main.

A way out is to compactify the extra dimension on an interval $I_1$ instead of a 
circle, which can support chiral fermions at the two ends of the interval. Since the interval
can be obtained easily from the circle by "orbifolding", i.e. by 
identifying points and fields in the circle theory under the 
$Z_2$ reflection operator ${\cal R}: x^5 \longrightarrow -x^5$, we will use 
the name orbifold when we refer to such a theory. A characteristic property of 
this orbifold is that it is defined on a space with two four dimensional boundaries
at each of the fixed points of the reflection action, where the gauge symmetry is reduced,
thus naturally differentiating the boundaries from the rest of the space, which we call the bulk.
An unfortunate consequence of field theories defined in such spaces is that the
all order finiteness of the Higgs mass arguments are not anymore applicable because of the
bulk-boundary interactions appearing at higher orders in perturbation theory which start to infect
the finite bulk mass with cut-off dependence \cite{vonGersdorff:2005ce}. 
This is not unexpected; it is known that handling non-renormalizable theories analytically 
is not easy, in fact there is no general prescription that can be used in these theories
such that their predictions are trustworthy.

We would therefore like here to start a systematic investigation of higher dimensional 
orbifold gauge theories from the point of view of a lattice regularization \cite{Irges:2004gy,Knechtli:2005dw,Irges:2006zf}.
The theory which will serve as our concrete example is an $SU(2)$ gauge theory which 
has the symmetry broken (by boundary conditions) to its $U(1)$ subgroup on the boundaries.
This, we will argue, is a promising way of approaching extra dimensional theories:
the goal is a non-perturbative understanding of a class of non-renormalizable theories and 
the hope is  a non-perturbative understanding of the SM Higgs mechanism. 

It turns out that phenomenology puts surprisingly tight constraints on models
and we will concentrate here on the two most immediate ones as one proceeds with the construction.
One is associated with the mere existence of a four dimensional effective action and the second,
closely related to SM phenomenology, the expected hierarchy of masses between the physical Higgs
and the massive gauge bosons in the broken phase of the SM. 

\subsection{Dimensional reduction}

The first issue is to show that an extra dimensional theory possesses in its parameter
space a regime where it undergoes some kind of effective dimensional reduction. 
This can happen, in principle, in more than one ways. Historically, the first mechanism of hiding
the extra dimension is going to the "Kaluza-Klein gauge" and making the 
size of the extra dimension small enough, so that its effects
are negligible up to a certain energy scale and at the same time keeping the couplings
perturbative so that the resulting effective theory can be useful for electroweak 
(or gravitational) physics.  
This is of course guaranteed as long as one treats the
radius of the circle $R$, the cut-off\footnote{
Extra-dimensional theories are non-renormalizable and make sense only with
a cut-off in place.} $\Lambda$
and the gauge coupling $g_5$ as independent parameters which is perhaps 
justified to a certain extent --
it depends essentially on how far one is willing to go in perturbation theory.
In a non-perturbative formulation on the other hand it becomes right away 
obvious that these three parameters are tightly connected
from the beginning and depending on where one sits in parameter space, a small change in one
of the parameters could result in a dramatic change of the system. 
From this point of view, the region in parameter space where the theory
describes the real world, (if it exists) could be only a small patch, which
makes one wonder whether the free stretching of the dimensionful parameters sometimes employed
to make a model phenomenologically viable is a valid operation.

More recently, mechanisms of dimensional reduction which depend on localization 
rather than compactification have been proposed. In the context of gauge theories,
all of these mechanisms involve one way or another a strong coupling and therefore
non-perturbative physics. One, is the so called layered phase originally
proposed in \cite{Fu:1983ei}. The idea is to investigate a five-dimensional
lattice model
where the gauge coupling in the four dimensional slices along the extra
dimension is different
from the gauge coupling that describes the interaction between the slices.
This anisotropy could give rise to a new, so-called layered, phase where the
static force is of Coulomb type in the four-dimensional slices and confining
along the extra dimension, thus providing a localization mechanism.
Some evidence for the existence of the layered phase in Abelian gauge theory
was recently given in \cite{Dimopoulos:2006qz}.
Another similar idea was due to \cite{Dvali:1996xe} where
it is assumed that the system has a phase where
the bulk is in a confined while
the boundary (defined by a domain wall) is in a deconfined phase,
which forces the boundary gauge fields to remain localized.
This idea was investigated on the lattice \cite{Laine:2004ji} and it was
found that the low-energy effective theory contains not only the localized
zero-modes but also higher Kaluza--Klein modes.

A different mechanism of dimensional reduction was proposed in the context
of the D-theory regularization of non-Abelian gauge theories
\cite{Chandrasekharan:1996ih,Brower:1997ha,Schlittgen:2000xg}.
Here a non-Abelian gauge theory in five dimensions arises as a low-energy
effective description of a five-dimensional quantum link model.
The size $L$ of four dimensions is taken to be infinity and the fifth
dimension has a finite size $R$. If the five-dimensional theory happens to
be in the Coulomb phase, the gluons are not massless but gain
a mass which is exponentially small in the size of the extra-dimension.
Thus by making $R$ larger the correlation length given by the inverse
gluon mass grows exponentially fast in $R$ and therefore the extra-dimension
disappears.

In this paper we will not be able to give a conclusive answer to this important 
issue since it would involve a very extensive scan of the parameter space.
We will give though some indirect evidence for dimensional reduction 
in the orbifold theory and we will leave the question of which mechanism 
is responsible for it, for a future work.
At a qualitative level, we will be able to map the part of the parameter space 
of our model where simulations were carried out onto the phase diagram of a 
known and well understood four dimensional theory, the Abelian Higgs model,
which provides our first indirect evidence for an effective dimensional reduction.  
Our more quantitative, but still indirect, evidence for 
dimensional reduction will be the measurement of the static potential between 
two infinitely heavy charged particles placed on 
four-dimensional slices located at the origin and in the middle 
of the fifth-dimension, separated by a distance $r$.
Using the fact that we measure a massive $U(1)$ gauge boson,
we will fit the numerical data for the static potential
to a five dimensional Yukawa potential and to a four dimensional Yukawa potential
and compare the fits in both cases.

\subsection{Hierarchy of masses}

A universal feature that perturbative five dimensional orbifold pure gauge theories 
seem to posses is that the gauge bosons that survive on the boundaries 
are all massless and the mass of the Higgs turns out to be too low.
In order to make some of the gauge bosons heavy one is forced to introduce 
fermions in the bulk \cite{Kubo:2001zc}, but still the ratio of the Higgs to the gauge boson mass is
${m_h}/{m_\g} << 1$ \cite{Scrucca:2003ra,Panico:2005dh}
and thus phenomenologically excluded, unless additional assumptions are made. 
This property has been traced to the fact that
since in five dimensions gauge invariance forbids a tree level Higgs mass, 
it has to be generated quantum mechanically. This one-loop mass
turns out to be suppressed compared to the gauge boson mass which is governed
by a vacuum expectation value of order one.
More concretely, at one loop, the Higgs mass is found to be $m_h\sim g_5/R^{3/2}$,
where $g_5$ is the five-dimensional gauge coupling, whereas  
the gauge boson mass is basically $m_\g \sim \a/R$ where $\a$ is a 
dimensionless vacuum expectation value obtained from minimizing the potential (see Appendix A).
Realistic models require a very small $\a$ even though typical one-loop
potentials tend to yield an $\a$ which is either zero or of order $\sim 0.1$. 

We will be able to compute the ratio $m_h/m_\g$ non-perturbatively and find 
that, contrary to perturbative expectations, a Higgs mechanism is at work,
resulting into a non-zero gauge boson mass, for a large part of the parameter space
(in fact for the whole range of parameters we were able to scan).  
Again, disentangling the precise dependence of this ratio on compactification and finite lattice size effects 
requires a larger scan of the parameter space, which will be the topic
of a future work.

\subsection{Organization of the paper}

In this paper we investigate numerically a one dimensional subspace of the 
parameter space of a five dimensional lattice gauge theory,
starting from the vicinity of a first order phase transition and approaching the perturbative 
domain. 
In section \ref{s_dimred}, using generic properties of the measured orbifold spectrum, 
we argue that the spectrum of the system in part of this region seems to 
be consistent with an effective dimensional
reduction. In section \ref{s_lat} we construct in detail the lattice regulated theory and
in particular its observables. 
In section \ref{s_numeric} we present in detail our 
quantitative evidence for spontaneous symmetry breaking and dimensional reduction. 
In section \ref{s_concl} we state our conclusions. 
Finally, for completeness we provide two detailed appendices with some background, appendix \ref{s_appa} on 1-loop results and \ref{s_appb} on the derivation
of the five-dimensional Yukawa potential.

%% file: sect2.tex
%%%%%%%%%%%%%%%%%%%%%%%%%%%%%%%%%%
\section{The orbifold effective theory \label{s_dimred}}
%%%%%%%%%%%%%%%%%%%%%%%%%%%%%%%%%%

We consider a pure $SU(2)$ Yang--Mills (YM) theory in five dimensions with gauge
potential $A_M$, $M=0,1,2,3,5$. The fifth dimension is an interval, obtained by 
identifying points and fields on a circle of radius $R$ under the
$Z_2$ reflection operator ${\cal R}: x^5 \longrightarrow -x^5$.
The projection breaks the gauge symmetry at the two ends of the interval according to
\bea
SU(2) & \longrightarrow & U(1) \,.
\eea
In the picture of dimensional reduction as it is known in finite temperature field
theory \cite{ZinnJustin-Book} the five-dimensional fields are expanded in a Fourier
series in the quantized momentum along the compact dimension. At low energies, in our
case much below the compactification scale given by the inverse radius $1/R$, we have
an effective four-dimensional theory of the zero-modes (i.e. constant along the fifth
dimension) of the fields. 

But the Fourier series that defines the Kaluza--Klein expansion breaks gauge invariance.
This we cannot afford at the non-perturbative level. Here the particle spectrum is read
off from correlations of five-dimensional, gauge invariant operators. These operators
are classified according to specific symmetries. If dimensional reduction occurs, a mass
gap in the various particle channels should be seen.
The ground state of the operators which corresponds to the
$A_5$ gauge field component\footnote{
In \sect{ss_higgsop} we will specify the gauge-invariant meaning of this, related to the
Polyakov line.}
has the quantum numbers of a complex scalar
and the ground state of the $A_{\m}, \m =0,1,2,3$ components\footnote{
In \sect{ss_photonop} we will construct a gauge-invariant operator that corresponds to the
gauge bosons, very much in analogy with the four-dimensional
$SU(2)$ Higgs model \cite{Montvay:1984wy}.}
have quantum numbers of a $U(1)$ gauge field from the point of view of four dimensions.
Clearly, the lowest lying spectrum of the orbifold theory should
coincide with the one of the four-dimensional Abelian Higgs model\footnote{
We thank U.-J. Wiese for his comments in this respect.},
if dimensional reduction works like in finite temperature field theory.
The spectra with the higher excitations included will be different, as in the
five-dimensional theory they are sensitive to the compactification scale $1/R$.
Before defining the orbifold theory in great detail and present extensive simulation results 
(which will be the topic of the next sections),
one would hope to be able to exploit this similarity of spectra
by mapping the phase diagram of the orbifold simulation on the phase diagram 
of the Abelian Higgs model.
Being able to do so, would be a first circumstantial evidence for an effective
dimensional reduction.

Next, we define the parameter space of our model.
In a lattice regularization, the cut-off is provided by the inverse lattice
spacing $\Lambda=a^{-1}$ and it preserves gauge invariance. That is why the lattice
formulation is particularly useful to study non-renormalizable theories, which
requires a cut-off.
The parameter space of the lattice theory is two dimensional\footnote{
For this discussion we assume the four-dimensional sizes of the lattice
to be infinite.} .
One of the parameters is the (dimensionless) number of lattice points in the fifth dimension
\bea
 N_5 & = & \frac{\pi R}{a} \label{Rlat}
\eea
and the second is the dimensionless five dimensional lattice coupling
\be
\b = \frac{2N}{g_5^2}a = \frac{2N}{g_0^2}\,, \label{betalat} 
\ee
where $g_0$ is the dimensionless bare gauge coupling.
For later reference, notice that in regions of the parameter space where $g_5$ has a mild cut-off
dependence, $\b$ is inversely proportional to $\Lambda$.
All observables are computed in units of the lattice spacing,
or equivalently in units of $R$, and are thus functions of $N_5$ and $\b$. 
One can also define a derived parameter, the coupling
\be
\beta_4^{eff}\equiv\frac{1}{g_4^2}  =  \frac{\pi R}{g_5^2},\label{b4}
\ee
for the effective four-dimensional theory.  
It is also possible in principle to use an anisotropic lattice \cite{Ejiri:2000fc} in which case
a lattice gauge coupling (and a different lattice spacing) is defined for the fifth dimension ($\b_5$)
and a different one for the four dimensional subspace ($\b_4$). 
Here we will restrict ourselves to isotropic lattices for which $\b_4=\b_5=\b$.
 
From simulations of the five dimensional $SU(2)$ gauge theory
on the orbifold $S^1/\mathbb{Z}_2$ with the number of points in the 
extra dimension fixed to $N_5=4$, 
and on an isotropic lattice with an associated coupling $\b$,
the spectrum can be safely determined when $\beta$ is larger
than a critical value $\beta_c=1.5975$, which
separates a confined ($\beta<\beta_c$) from a deconfined ($\beta>\beta_c$)
phase. The spectrum measured in simulations corresponding to the deconfined phase consists of
a massive Higgs ground state and a massive Abelian gauge boson, along with their
excitations \cite{Irges:2006zf} which implies that our system in
this region of parameter space crosses from a confined phase into a Higgs phase.
\footnote{If the four-dimensional effective theory were just a pure $U(1)$ gauge theory,
(due to the orbifold breaking of the $SU(2)$ symmetry) and dimensional reduction occurs,
one should be able to map our results onto the four-dimensional Abelian model.
For $N_5=4$ we have $\beta_4^{eff}=\beta$. The pure compact $U(1)$ gauge theory
has a phase transition at $\beta_c=1.01$ \cite{Arnold:2002jk} and therefore we would end up
in its Coulomb phase. Clearly, this can not be sufficient to describe our system
because, as already mentioned, in the deconfined phase we measure
not only massive scalars but also a massive gauge boson.}
In order to be able to make a comparison with the Abelian Higgs model
we now recall a few well known facts.

We summarize results from analytic \cite{Fradkin:1978dv,Osterwalder:1977pc} and
numerical \cite{Jansen:1985cq} studies of the four dimensional Abelian Higgs model.
The Euclidean action may be written as
\bea
 S & = & \sum_x\Bigg\{\lambda(\phi^\dagger(x)\phi(x)-1)^2+\phi^\dagger(x)\phi(x)
 \nonumber \\
 & &
 -\kappa\sum_\mu\left(\phi^\dagger(x){\rm e}^{iqA_\mu(x)}\phi(x+a\hat{\mu}) + {\rm c.c.}\right)
 \Bigg\} - \frac{\beta}{2}\sum_pU_p \,. \label{action}
\eea
Here, $\phi(x)$ is the complex Higgs field, $q$ is its charge and $\kappa$ the hopping
parameter related to the bare mass $m_0$ through
\bea
 a^2m_0^2 & = & \frac{1-2\lambda}{\kappa}-8 \,.
\eea
The analytic study in
\cite{Fradkin:1978dv} was performed in the limit $\lambda\to\infty$ where the length of the Higgs
field $\rho$ is frozen to 1. In the unitary gauge the action then reads
\bea
 S & = & -\kappa\sum_\mu2\cos[qA_\mu(x)]
 - \frac{\beta}{2}\sum_pU_p \label{actionunitaryfrozen} \,.
\eea
There are two interesting cases. One is when the Higgs field is in the fundamental
representation, that is it has charge $q=1$ and the second is when 
it has charge $q=2$. For us it is the second case that is relevant since,
as will show in the next section, the lattice operator which is identified with
the Higgs has charge two.
 
%
%%%%%%%%%%%%%%%%%%%%%%%%%%%%%%%%%%%%%%%%%%%%%%%%%%%%%%%%%%%%%%%%%%%%%%%%%%%%%
\begin{figure}[t]
\centerline{\epsfig{file=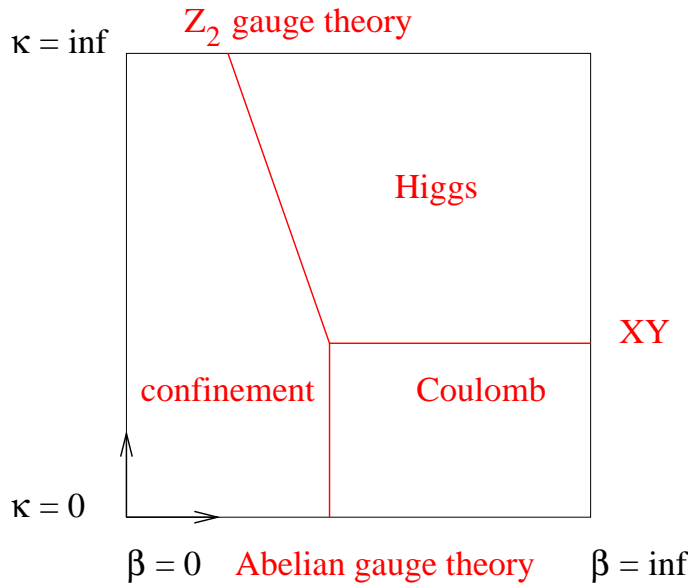,width=9cm}}
\caption{Phase diagram of the Abelian Higgs model for the Higgs field
with charge $q=2$. From studies at $\lambda=\infty$ \cite{Fradkin:1978dv}.
\label{f_phq2}}
\end{figure}
%%%%%%%%%%%%%%%%%%%%%%%%%%%%%%%%%%%%%%
%

%\begin{figure}[!b]
%  \begin{center}
%    \includegraphics[width=4in]{4DAbHiggs2.pdf}
%  \end{center}
%   \caption{\small Phase diagram of the Abelian Higgs model for the Higgs field
%with charge $q=2$. From studies at $\lambda=\infty$ \cite{Fradkin:1978dv}.  }
%\label{f_phq2}
%\end{figure}

We inspect
the action for $\lambda=\infty$ in the unitary gauge, \eq{actionunitaryfrozen},
in the limit $\kappa\to\infty$. In this limit the gauge variable $A_\mu(x)$ can
take, for $q=2$, two values, 0 or $\pi$, corresponding to $\mathbb{Z}_2$ gauge links
$U(x,\mu)=\pm1$. (For general charge $q$ it will be a $\mathbb{Z}_q$ gauge theory).
The $\mathbb{Z}_2$ gauge theory has a second order phase transition.
The analytic study of \cite{Fradkin:1978dv} in the $\lambda=\infty$ case concludes
that there are three distinct phases, sketched in \fig{f_phq2}. The main difference
compared to the $q=1$ case is a phase boundary that separates the 
Higgs from the confinement phase. The
static potential in the Higgs phase is of Yukawa type. In the confinement phase
it rises linearly (area law for Wilson loops) which is not the case in the
confinement region for a fundamental Higgs.  

Clearly, this is the version of the Abelian Higgs model 
on which the orbifold theory can be naturally mapped
since the lowest lying spectra are the same and the part of the orbifold phase
diagram we have investigated can be recognized inside the 
phase diagram of the $q=2$ Abelian Higgs model.

%% file: sect3.tex
%%%%%%%%%%%%%%%%%%%%%%%%%%%%%
\section{Orbifold on the lattice \label{s_lat}}
%%%%%%%%%%%%%%%%%%%%%%%%%%%%%

Gauge theories on the orbifold can be discretized on the lattice
\cite{Irges:2004gy,Knechtli:2005dw}.
One starts with a gauge theory formulated on a five-dimensional torus with
lattice spacing $a$ and periodic boundary conditions in all directions 
$M=0,1,2,3,5$.
The spatial directions ($M=1,2,3$) have length $L$,
the time-like direction ($M=0$) has length $T$, and the
extra dimension ($M=5$) has length $2\pi R$. The coordinates of the 
points are
labelled by integers $n\equiv\{n_M\}$ and the gauge field is the set of link
variables $\{U(n,M)\in SU(N)\}$.
The latter are related to a gauge potential 
$A_M$ in the Lie algebra of $SU(N)$
by $U(n,M) = \exp\{a A_M(n)\}$.
Embedding the orbifold action in the gauge field
on the lattice amounts to imposing on the links the $\mathbb{Z}_2$ 
projection
\be
(1-\G)U(n,M) = 0 \,, \label{orbiproj}
\ee
where $\G = {\cal R} {\cal T}_g$. Here, ${\cal R}$ is the
reflection operator that acts as ${\cal R}\,n=(n_\mu,-n_5)\equiv{\bar n}\;(\mu=0,1,2,3)$
on the lattice and as ${\cal R}\,U(n,\m) = U ({\bar n},\m)$
and ${\cal R}\,U(n,5) = U^\dagger({\bar n}-{\hat 5},5)$ on the links.
The group conjugation ${\cal T}_g$ acts only on the links, as
${\cal T}_g U(n,M) = g U(n,M) g^{-1}$,
where $g$ is a constant $SU(N)$ matrix with the property that
$g^2$ is an element of the centre of $SU(N)$.
For $SU(2)$ we will take $g=-i\s^3$. Only gauge transformations 
$\{\Omega(n)\}$
satisfying $(1-\G)\Omega=0$ are consistent with \eq{orbiproj}. This 
means that
at the orbifold fixed points, for which $n_5=0$ or $n_5=N_5$, the
gauge group is broken to the subgroup that commutes with $g$. For 
$SU(2)$ this
is the $U(1)$ subgroup parametrized by $\exp(i\phi\s^3)$, where $\phi$ 
are compact
phases.

After the projection in \eq{orbiproj},
the fundamental domain is the strip
$I_1 = \{n_\m , 0\le n_5 \le N_5\}$.
The gauge-field action on $I_1$ is taken to be the Wilson action
\be
S_W^{\rm orb}[U] = \frac{\b}{2N} \sum_p w(p)\; {\tr}\, \{1-U(p)\}, 
\label{wila}
\ee
where the sum runs over all oriented plaquettes $U(p)$ in $I_1$.
The weight $w(p)$ is 1/2 if $p$ is a plaquette in the $(\m\n)$ planes
at $n_5=0$ and $n_5=N_5$, and 1 in all other cases.
Dirichlet boundary conditions are imposed on the gauge links
\bea
 U(n,\mu) & = & g\,U(n,\mu)\,g^{-1} \quad
 \mbox{at $n_5=0$ and $n_5=N_5$.} \label{Dbclat}
\eea
The gauge variables at the boundaries are not fixed but are
restricted to the subgroup of $SU(N)$, invariant under ${\cal T}_g$.
The Wilson action together with these boundary conditions reproduce the 
correct
naive continuum gauge action and boundary conditions on the components 
of the
five-dimensional gauge potential \cite{Irges:2004gy}.
For example, for $SU(2)$, $A_\m^{3}$ (``photon'')
and $A_5^{1,2}$ (``Higgs'') satisfy Neumann boundary conditions and
$A_\m^{1,2}$ and $A_5^{3}$ Dirichlet ones.

One of the main results of \cite{Irges:2004gy} was to show, through a
geometrical construction, that the orbifold projection equation \eq{orbiproj}
implies the absence of a boundary counterterm for the Higgs mass.
Given the explicit breaking of the gauge invariance at the boundaries, a
boundary mass term for $A_5$ is invariant under the unbroken gauge group.
In the continuum this mass term would be $\tr\{[A_5,g][A_5,g^{-1}]\}$
evaluated at the boundaries $x_5=0$ and $x_5=\pi R$. If present, such a term
would imply a quadratic sensitivity of the Higgs mass to the cut-off. For
the lattice action \eq{wila} this would require to add a boundary action
term with an additional coefficient $\tilde{\mu}$. As the lattice spacing $a$
changes, a fine tuning of $\tilde{\mu}$ would be required to keep the Higgs
mass finite. Fortunately this term is absent and the orbifold action is simply
\eq{wila}. Since five-dimensional theories are non-renormalizable, they make
sense only as effective theories for energy scales much below the cut-off.
On the lattice this is the Symanzik effective action
\cite{Symanzik:1981hc,Symanzik:1983dc,Symanzik:1983gh,Luscher:1998pe},
a continuum action which is a systematic expansion in the lattice spacing $a$.
For the orbifold theory the Symanzik effective action is \cite{Irges:2004gy}
\bea
S & = & -\frac{1}{2g_5^2}\Bigg[\int{\rm d}^5z\,\tr\{F_{MN}F_{MN}\} 
\nonumber \\
& & +ab_1\int_{z=\{0,\pi R\}}{\rm d}^4x\,{\rm Re}\,\tr\{gF_{MN}F_{MN}\}
    +ab_2\int_{z=\{0,\pi R\}}{\rm d}^4x\,{\rm Re}\,\tr\{gF_{MN}gF_{MN}\}
\nonumber \\
& & +a^2c\int{\rm d}^5z\,\tr\{D_LF_{MN}D_LF_{MN}\} + \ldots \Bigg] \,,
\eea
where $F_{MN}$ is the field strength tensor, $D_L$ its covariant derivative
and the coefficients $b_1,b_2,c,\ldots$ are computable in perturbation theory.
At 1-loop, $b_1\neq0$ and $b_2=0$ \cite{vonGersdorff:2002as}. A boundary
mass term for the Higgs would appear with a coefficient $\tilde{\mu}/a^2$.
 
%
%%%%%%%%%%%%%%%%%%%%%%%%%%%%%%%%%%%%%%
\subsection{Operators for the Higgs}
%%%%%%%%%%%%%%%%%%%%%%%%%%%%%%%%%%%%%%
%

If the fifth dimension were infinite, the gauge links $U(n,5)$ would be
gauge-equivalent to the identity, which corresponds to the continuum
axial gauge $A_5\equiv0$.
On the circle $S^1$ one can gauge-transform
$U(n,5)$ to an $n_5$-independent matrix $V(n_\mu)$ that
satisfies $P=V^{2N_5}$, where $P=P(n_\mu)$
is the Polyakov line winding around the extra dimension at
four-dimensional location $n_\mu$.
Therefore an extra-dimensional potential
$\A5$ can be defined on the lattice, through $V=\exp\{a\A5\}$, as
\be
 a\A5 = \frac{1}{4N_5}(P-P^\dagger) + {\rm O}(a^3) \,. \label{A5lat}
\ee
At finite lattice spacing the O($a^3$) corrections in \eq{A5lat}
are neglected. One easily checks that ${\cal R}\,P \,=\, P^{\dagger}$ and so
${\cal R}\,(A_5)_{\rm lat}=-(A_5)_{\rm lat}$, as it should be to have the
same transformation behaviour as $A_5$ in the continuum.

%\begin{figure}[!t]
%  \begin{center}
%    \includegraphics[width=3.5in]{higgs.pdf}
%  \end{center}
%   \caption{\small The Polyakov line.}
%  \label{f_higgs}
%\end{figure}

%
%%%%%%%%%%%%%%%%%%%%%%%%%%%%%%%%%%%%%%%%%%%%%%%%%%%%%%%%%%%%%%%%%%%%%%%%%%%%%
\begin{figure}[!t]
\centerline{\epsfig{file=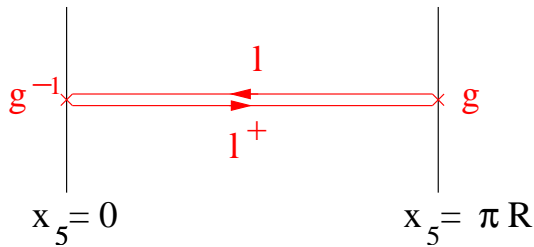,width=7cm}}
\caption{The Polyakov line $P$ on $S^1/\mathbb{Z}_2$.
\label{f_higgs}}
\end{figure}
%%%%%%%%%%%%%%%%%%%%%%%%%%%%%%%%%%%%%%%%%%%%%%%%%%%%%%%%%%%%%%%%%%%%%%%%%%%%%
%

In order to construct the gauge potential $A_5$ on the orbifold
$S^1/\mathbb{Z}_2$
we start from the circle $S^1$ parametrized by the coordinates
$n_5=-N_5,\ldots,N_5-1$ ($N_5=\pi R/a$ is identified with $-N_5$).
We impose on the links building the Polyakov line $P$
\bea
 P(n_\mu) \,=\, U((n_\mu,0),5)\,\ldots\,U((n_\mu,N_5-1),5)\,U((n_\mu,-N_5),5)
                \,\ldots\,U((n_\mu,-1),5)
\eea
the orbifold projection \eq{orbiproj}. The result is
\bea
 P(n_\mu) & = & l(n_\mu)\,g\,l^\dagger(n_\mu)\,g^{-1} \,,\label{Lorbi}
\eea
with
\bea
 l(n_\mu) & = & U((n_\mu,0),5)\,U((n_\mu,1),5)\,\ldots\,U((n_\mu,N_5-1),5)
 \,. \label{linpoly}
\eea
The Polyakov line \eq{Lorbi} on $S^1/\mathbb{Z}_2$ is
shown schematically in \fig{f_higgs} and from it we
define $\A5(n_\mu)=\{P(n_\mu)-P^\dagger(n_\mu)\}/(4N_5)$.
Being anti-Hermitian, the field $\A5$ can be represented using the
unit matrix and the Hermitian generators $T^A$ of $SU(N)$:
\bea
(A_5)_{\rm lat} & = & -ig_0(A_5^0\,\mathbf{1}_N + A_5^A\,T^A) \,.
\eea
In order to construct the Higgs field on $S^1/\mathbb{Z}_2$
we have to project\footnote{
In the continuum the orbifold projection selects automatically the components
$A_5^{\hat{a}}$. On the lattice, due to the lattice artifacts $O(a^3)$ in the
definition \eq{A5lat}, ``wrong'' components $(A_5)_{\rm lat}^a$ can be non-zero.}
$(A_5)_{\rm lat}$
onto the components $A_5^{\hat{a}}$ for which $g\,T^{\hat{a}}\,g^{-1}=-T^{\hat{a}}$
(the other generators have $g\,T^a\,g^{-1}=T^a$).
The Higgs field is now defined as in the continuum
\bea
 \Phi(n_\mu) & = & [a(A_5)_{\rm lat}(n_\mu),g] \,=\,
 2iag_0A_5^{\hat{a}}(n_\mu)\,g\,T^{\hat{a}} \,.
 \label{higgsfield}
\eea
Note that the commutator projects out the identity component of $(A_5)_{\rm lat}$.
Under gauge transformation $\Omega(n)$,
$l(n_\mu)\longrightarrow\Omega(n_\mu,0)\,l(n_\mu)\,\Omega(n_\mu,N_5)^{-1}$.
Since $[\Omega(n),g]=0$ at the orbifold boundaries $n_5=0$ and $n_5=N_5$,
it follows that
\bea
\Phi(n_\mu) & \longrightarrow & \Omega(n_\mu,0)\,\Phi(n_\mu)\,
\Omega(n_\mu,0)^{-1} \,. \label{gthiggs}
\eea
The Higgs field $\Phi$ transforms like a field strength tensor at the
boundary. In the special case of gauge group $SU(2)$ only a $U(1)$ gauge
symmetry survives at the boundaries.
If we parameterize the $U(1)$ boundary gauge transformations
by $\Omega(n_\mu,0)=\exp\{i\omega(n_\mu)\s^3\}$, the Higgs field transforms as
\bea
 \Phi\,=\,
 \left(\begin{array}{cc}
 0 & h=\phi_1-i\phi_2 \\
 h^\dagger=\phi_1+i\phi_2 & 0
 \end{array}\right) 
 & \longrightarrow &
 \left(\begin{array}{cc}
 0 & {\rm e}^{2i\omega}h \\
 {\rm e}^{-2i\omega}h^\dagger & 0
 \end{array}\right) \,. \label{charge2}
\eea
showing that it has charge 2 under the $U(1)$ gauge group.

Since $\tr\{\Phi\}=0$, in order to extract the Higgs mass we define
\bea
 H(n_0) & = & \left(\frac{a}{L}\right)^3
 \sum_{n_1,n_2,n_3} \tr\{\Phi(n_\mu)\,\Phi^\dagger(n_\mu)\}
 \label{Hfield}
\eea
and build the connected correlation
\bea
 C(t) & = &
 \frac{a}{T}\sum_{n_0}\left\{\langle H(n_0)\,H(n_0+t/a) \rangle
 -\langle H(n_0) \rangle\,\langle H(n_0+t/a) \rangle  \right\} 
 \label{conncorr} \\
 & \stackrel{t\to\infty}{\longrightarrow} & {\rm const.}\times{\rm e}^{-m_h t} 
 \,, \label{expconncorr} \nonumber
\eea
and the Higgs mass $m_h$ can be extracted from the effective masses
$am_{h,{\rm eff}}(t+a/2)\,=\,\ln\{C(t)/C(t+a)\}$.
Writing the correlation $C(t)$ as
\bea
C(t) & = & \langle\left( H(t) - \langle H\rangle \right) 
                  \left( H(0) - \langle H\rangle \right)\rangle
\eea
one can see that it is a sum of positive and negative numbers of order
one, the result being a small number due to cancellations. 
On the other hand, the variation 
\bea
\D C & = &  \langle\left( H(t) - \langle H\rangle\right)^2 
                   \left( H(0) - \langle H\rangle\right)^2 \rangle - C^2
\eea
is a sum of positive numbers of order one minus a very small number
and hence of order one. 
Furthermore, $\D C$ is essentially independent of $t$.
This means that the error in the effective Higgs masses is approximately
\bea
\D am_{h,{\rm eff}} & \simeq &
\D C \sqrt{\left(\frac{1}{C(t)}\right)^2 + \left(\frac{1}{C(t+a)}\right)^2} \,,
\eea
and, since $C(t) \sim e^{-m_h t}$, its $t$--dependence is 
\bea
\D am_{h,{\rm eff}} & \sim & e^{m_h t}.\label{emerror}
\eea
In the above, $m_h$ is the plateau value of the Higgs mass which is
constant and therefore one expects that the error in $m_h$
increases with $t$ exponentially.

%
%%%%%%%%%%%%%%%%%%%%%%%%%%%%%%%%%%%%%%%%%%%%%%%%%
\subsubsection{Variational technique}
%%%%%%%%%%%%%%%%%%%%%%%%%%%%%%%%%%%%%%%%%%%%%%%%%
%

It turns out that the correlation function \eq{conncorr} suffers from a loss 
of significance
in numerical simulation. This is the, unfortunately, common problem of signals,
which are exponentially small in the time $t$, with an almost constant variance, and
hence constant statistical error. The ratio of the signal to the error falls off
exponentially in the time $t$;
in the large-$t$ region, where the leading exponential decay \eq{expconncorr} due
to the Higgs mass should be seen, the signal is lost. This is reflected in the
exponentially growing error of the effective masses \eq{emerror}.

To cure this problem we employ the variational technique of
\cite{Luscher:1990ck}.
A basis of Higgs operators is constructed and the best operator
to capture the Higgs mass in \eq{expconncorr},
i.e. whose overlap with the eigenstate of the Hamiltonian is the largest,
will be a linear combination of these basis fields.
If the contributions from excited states are
suppressed then the leading exponential \eq{expconncorr}
can be extracted at smaller values of $t$, where the signal might not be lost.

Here we sketch how this works;
more details can be found in \cite{Luscher:1990ck}. We construct a set
of Euclidean fields $\mO_i$, $i=1,\ldots,r$ with the same quantum numbers as
$H\equiv\mO_1$ in \eq{Hfield}. Then we build the matrix correlation function
\bea
C_{ij}(t) & = & \langle\mO_i(t)\mO_j(0)^*\rangle -
             \langle\mO_i(t)\rangle\,\langle\mO_j(0)^*\rangle \,, \label{matconcorr}
\eea
and write the spectral decompositions
\bea
\langle\mO_i(t)\mO_j(0)^*\rangle & = &
\frac{1}{Z}\sum_{m,n}{\rm e}^{-E_nT-t(E_m-E_n)}A_{nm}^{(i)}A_{nm}^{(j)\,*} \,
\label{specdeccor} \\
\langle\mO_i(t)\rangle & = & \frac{1}{Z}\sum_n{\rm e}^{-E_nT}A_{nn}^{(i)} \,,
\label{specdecZ}
\eea
where
\bea
 Z\,=\,\sum_n{\rm e}^{-E_nT} & \mbox{and} & 
 A_{mn}^{(i)}\,=\,\langle m|\mO_i(0)|n\rangle \,.
\eea
Here $m,n=0,1,2,...$
label the eigenstates $|m\rangle$ with energy eigenvalue $E_m$ of
the Hamiltonian $\mathbb{H}$ and $T$ is the temporal size of the lattice.
We use the same symbol $\mO_i$ to denote the
Euclidean field and the corresponding operator in the Hamiltonian formulation.

There are two effects, which derive from the finiteness of $T$ and the periodic
boundary conditions in time (which imply taking the overall trace in the spectral
decomposition) \cite{Montvay:1987us}. Firstly,
if the operators $\mO_i$ have a non-vanishing expectation value the
connected correlation functions \eq{matconcorr} have in general t-independent
contributions
\bea
 {\rm e}^{-m_hT}\,\left[A_{00}^{(i)}-A_{11}^{(i)}\right]\,
 \left[A_{00}^{(j)}-A_{11}^{(j)}\right]^* \,, \label{tindep}
\eea
where $m_h=E_1-E_0$ is the mass gap, i.e. the Higgs ground state mass.

Secondly, in the limit that $T$, $t$ are both large and the difference $T-t$
is close to $t$, the matrix correlation function \eq{matconcorr}
has the leading behavior
\bea
 C_{ij}(t) & \longrightarrow & \sum_{n>0}\left[{\rm e}^{-(E_n-E_0)t}+
 {\rm e}^{-(E_n-E_0)(T-t)}\right] A_{n0}^{(i)}A_{n0}^{(j)\,*} + \nonumber \\
 & & \sum_{m,n>0}{\rm e}^{-(E_n-E_0)T-(E_m-E_n)t}A_{nm}^{(i)}A_{nm}^{(j)\,*} \,. \label{Casymptotic}
\eea
Here we have assumed that $A_{mn}=A_{nm}^*$ and $C_{ij}(t)$ is real,
which is true if the operators $\mO_i$ are Hermitian.
If we take $t=T/2$, the contribution of the second term in \eq{Casymptotic}
goes like $\exp\{-[(E_n+E_m)/2-E_0]T\}$ compared to $\exp\{-(E_n-E_0)T/2\}$ of the first
term. The second term is hence subleading and can be neglected. We get
\bea
 C_{ij}(t) & \longrightarrow & C_{ij}^\prime(t) + C_{ij}^\prime(T-t) \,,\quad
 C_{ij}^\prime(t) = \sum_{n>0}A_{n0}^{(i)}A_{n0}^{(j)\,*}
 {\rm e}^{-(E_n-E_0)t} \,. \label{Csymmetric}
\eea
The correlation function is a sum of two contributions and is symmetric
about $t=T/2$.

First we assume that $T$ is large enough so that the t-independent contribution
\eq{tindep} is negligible and that $t$ is small enough
so that the contribution $T-t$ is also negligible. Then the correlations
\eq{matconcorr} have the $T=\infty$ behavior\footnote{
Then in \eq{specdeccor} and \eq{specdecZ} only the term $n=0$ contributes.}
(assumed in \cite{Luscher:1990ck})
\bea
C_{ij}(t) & = &
\sum_{\alpha=1}^{\infty}A_{\alpha 0}^{(i)}A_{\alpha 0}^{(j)*}{\rm e}^{-tW_\alpha}
\,,\quad
W_{\alpha}=E_\alpha-E_0 \,.
\eea
The lowest masses $W_\alpha$ can be extracted by solving the generalized eigenvalue
problem
\bea
C(t)_{ij}\psi_{\alpha,j}(t,t_0) & = & 
\lambda_\alpha(t,t_0)C_{ij}(t_0)\psi_{\alpha,j}(t,t_0) \,, \label{geneigprob}
\eea
where the correlation matrix is taken at variable time $t$ and at fixed time $t_0$
(we may set $t_0=0$). From a mathematical lemma proved in \cite{Luscher:1990ck} it follows
that
\bea
\lambda_\alpha(t,t_0) & \stackrel{t\to\infty}{=}
c_\alpha{\rm e}^{-tW_\alpha}\left[1+{\rm O}(e^{-t\Delta W_\alpha})\right] \,,
\quad \alpha=1,\ldots,r \,, \label{geneig}
\eea
where $c_\alpha>0$ and $\Delta W_\alpha = \min_{\beta\neq\alpha}|W_\alpha-W_\beta|$.
One expects that $c_\alpha\simeq{\rm e}^{t_0W_\alpha}$ and that the
coefficients of the correction terms in \eq{geneig}, because of the excited states,
are suppressed. The masses can be extracted from
\bea
aW_{\alpha}(t+a/2) & = & \ln\left(\frac{\lambda_\alpha(t,t_0)}{\lambda_\alpha(t+a,t_0)}
\right) \label{geneigmasses}
\eea
at moderately large value of $t$.

If the contribution $T-t$ is not negligible, as it happens when $t$ approaches
$T/2$ (and we might be forced to go to such large values of $t$ to find a plateau),
then a different formula has to be used. The starting point is \eq{Csymmetric}.
If we say that $\lambda_\alpha^\prime(t,t_0)$ solves the generalized eigenvalue problem
\eq{geneigprob} for the matrix $C^\prime$ defined in \eq{Csymmetric}, then it is
straightforward to show that
\bea
 \lambda_\alpha(t,t_0) & = & \lambda_\alpha^\prime(t,t_0) +
                             \lambda_\alpha^\prime(T-t,t_0)
\eea
solves the generalized eigenvalue problem for the full matrix $C$. Using \eq{geneig}
for $\lambda^\prime$ we get the formula
\bea
 \lambda_\alpha(t,t_0) & = & 2c_\alpha{\rm e}^{-W_\alpha T/2}\cosh[(T/2-t)W_\alpha] \,.
\eea
By using the ratio $r_{12}=\lambda_\alpha(t_1,t_0)/\lambda_\alpha(t_2,t_0)$
and the definitions
\bea
 x={\rm e}^{-W_\alpha} \,, & \tau_1=T/2-t_1 \,, & \tau_2=T/2-t_2 \,,
\eea
it is easy to arrive at the equation
\bea
 r_{12}\left(x^{\tau_2}+x^{-\tau_2}\right) - \left(x^{\tau_1}+x^{-\tau_1}\right)
 & = & 0
\eea
to be solved numerically (Newton-Raphson) for $x$. We take $t_0=0$, 
$t_1=t$ and $t_2=t+a$ for $t=a,2a,\ldots,T/2-a$.

%
%%%%%%%%%%%%%%%%%%%%%%%%%%%%%%%%%%%%%%%%%%%%%%%%%
\subsubsection{Higgs operators \label{ss_higgsop}}
%%%%%%%%%%%%%%%%%%%%%%%%%%%%%%%%%%%%%%%%%%%%%%%%%
%

%\begin{figure}[!b]
%  \begin{center}
%    \includegraphics[width=3.5in]{higgsdisp.pdf}
%  \end{center}
% \caption{\small The displaced Polyakov line }
%  \label{f_higgsdisp}
%\end{figure}

%
%%%%%%%%%%%%%%%%%%%%%%%%%%%%%%%%%%%%%%%%%%%%%%%%%%%%%%%%%%%%%%%%%%%%%%%%%%%%%
\begin{figure}[!b]
\centerline{\epsfig{file=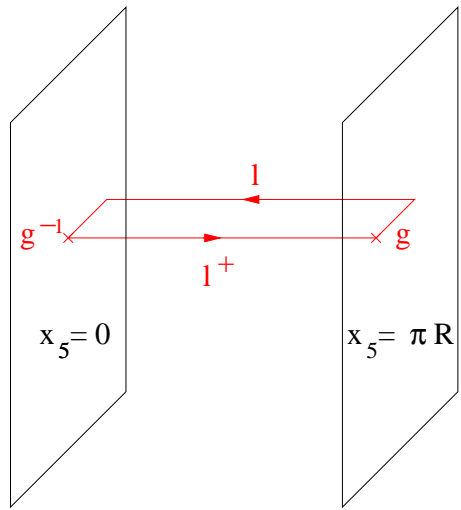,width=6cm}}
\caption{The displaced Polyakov line $L$ on $S^1/\mathbb{Z}_2$.
\label{f_higgsdisp}}
\end{figure}
%%%%%%%%%%%%%%%%%%%%%%%%%%%%%%%%%%%%%%%%%%%%%%%%%%%%%%%%%%%%%%%%%%%%%%%%%%%%%
%

A basis of Higgs operators $\mO_i=H_i$, defined as in \eq{Hfield},
can be constructed by modifying the definition of the Higgs field $\Phi$
to create a set of fields $\Phi_i$.
We can for example consider displaced Polyakov lines on $S^1/\mathbb{Z}_2$
as shown in \fig{f_higgsdisp}. The position where the displacement in one
of the spatial directions $k=1,2,3$ takes place can be varied along the
extra coordinate $n_5=0,1,\ldots,\lfloor N_5/2\rfloor$. The
displacements at the other $n_5$ values are equivalent by symmetry with
respect to reflection $n_5\longrightarrow N_5-n_5$.
For the displacement at $n_5=0$ \eq{Lorbi} is replaced by
\bea
 P^{(0)}(n_\mu) & = & 
 \frac{1}{6}\sum_{\stackrel{n_0=n^\prime_0}{|n-n^\prime|=a}}
 U(n_\mu,n^\prime_\mu)|_{n_5=0}\,l(n^\prime_\mu)\,
 U^\dagger(n_\mu,n^\prime_\mu)|_{n_5=N_5}\,g\,l^\dagger(n_\mu)\,g^{-1}
 \label{displacepoly} \,,
\eea
and similar expressions $P^{(n_5)}$ for $n_5=1,\ldots,\lfloor N_5/2\rfloor$.
In \eq{displacepoly}
$U(n_\mu,n_\mu^\prime)$ is the parallel transporter from the four-dimensional
point $n_\mu^\prime$ to $n_\mu$ in the indicated slice $n_5$ in the extra dimension.

%\begin{figure}[!b]
%  \begin{center}
%    \includegraphics[width=3.5in]{higgssmear.pdf}
%  \end{center}
%   \caption{\small A smearing procedure for the Higgs field.}
%  \label{f_higgssmear}
%\end{figure}

%
%%%%%%%%%%%%%%%%%%%%%%%%%%%%%%%%%%%%%%%%%%%%%%%%%%%%%%%%%%%%%%%%%%%%%%%%%%%%%
\begin{figure}[t]
\centerline{\epsfig{file=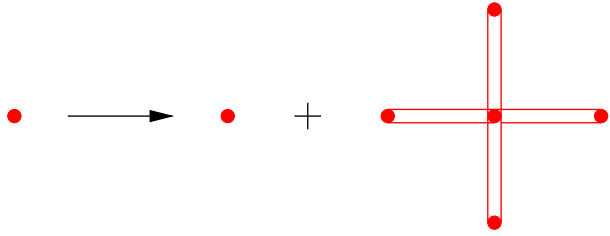,width=8cm}}
\caption{A smearing procedure for the Higgs field $\Phi$.
\label{f_higgssmear}}
\end{figure}
%%%%%%%%%%%%%%%%%%%%%%%%%%%%%%%%%%%%%%%%%%%%%%%%%%%%%%%%%%%%%%%%%%%%%%%%%%%%%
%

Yet another possibility to create fields for the variational basis is to consider
smeared Higgs fields. The simplest is shown in \fig{f_higgssmear}: the field
$\Phi(n_\mu)$ (represented by a thick point) is replaced by a
smeared field $\Phi^\prime(n_\mu)$ made of a linear combination of $\Phi(n_\mu)$ and the
nearest neighbour fields in the three dimensional space
\bea
 \Phi^\prime(n_\mu) & = & 
 (1-\alpha)\,\Phi(n_\mu) + 
 \frac{\alpha}{6}\sum_{\stackrel{n_0=n^\prime_0}{|n-n^\prime|=a}}
 U(n_\mu,n_\mu^\prime)|_{n_5=0}\,\Phi(n_\mu^\prime)\,
 U^{\dagger}(n_\mu,n_\mu^\prime)|_{n_5=0} \,. \label{smear1}
\eea
The smearing parameter is $\alpha$.
The definition of \eq{smear1} ensures that $\Phi^\prime$ transforms as $\Phi$ under
the gauge transformation in \eq{gthiggs}. Variants of the smearing technique can be found
in \cite{Knechtli:1999tw}. The smearing procedure \eq{smear1} can be iterated a number of times, which we take to be 3. We set $\alpha=0.7$.

Finally, the gauge links used to construct all the Higgs fields $\Phi_i$
are replaced by smeared gauge links. This is a very simple
procedure to create extended operators. We have implemented
APE smearing \cite{Albanese:1987ds}
for the spatial links, i.e. $U(z,k)$, $k=1,2,3$ and
$U(z,5)$. The links are decorated with staples in the spatial directions $l=1,2,3$, clearly with the restriction $l\neq k$ for $U(z,k)$.
The smeared link $U^\prime$ is obtained by adding to $(1-\alpha)U$ the
sum of the decorating staples multiplied by $\alpha$/(number of staples),
where $\alpha$ is the smearing parameter.
The smeared link is projected back onto $SU(2)$.
The smearing of the gauge links is iterated 3 times with $\alpha=0.75$.

%
%%%%%%%%%%%%%%%%%%%%%%%%%%%%%%%%%%%%%%%%%%%%%%%%%
\subsection{Photon operators \label{ss_photonop}}
%%%%%%%%%%%%%%%%%%%%%%%%%%%%%%%%%%%%%%%%%%%%%%%%%
%

In this section we describe the construction of operators in the
$SU(2)$ orbifold, which create the gauge boson (photon) associated
with the unbroken $U(1)$ gauge group on the boundaries.
For each Higgs field $\Phi_i$ constructed as described in \sect{ss_higgsop}
we define the $SU(2)$ valued quantity
(in the following we suppress the index $i$)
\bea
\a (n_\mu) & = & \frac{\Phi(n_\mu)}{\sqrt{\det(\Phi(n_\mu))}} \,.
\eea
We note that from \eq{higgsfield} and $g^\dagger=-g$ (which holds for SU(2))
it follows $\a^\dagger=-\a$.
For the three spatial directions $k=1,2,3$ we define a ``decorated double link''
\bea
V(n_\mu,k) & = & U(n_\mu,k)\,\a(n_\mu+a {\hat k})\,U^\dagger(n_\mu,k)\,\a(n_\mu) \quad
\mbox{(no sum over $k$)} \,,
\eea
which transforms like $\Phi$ under gauge transformations.
In analogy with the definition \cite{Montvay:1984wy} in the standard $SU(2)$ Higgs model,
we define an operator for the photon field
\bea
W_k(n_\mu) & = & -i\tr\{\s^3\,V(n_\mu,k)\} \,. \label{photonfieldx}
\eea
We will consider this field projected to zero three-dimensional momentum
\bea
W_k(n_0) & = & \left(\frac{a}{L}\right)^3\sum_{n_1,n_2,n_3}W_k(n_\mu) \,.
\label{photonfield}
\eea
It is not difficult to show that the operator $W_k(n_0)$ is real,
invariant under the group conjugation ${\cal T}_g$ and odd under
the three-dimensional parity transformation.
In order to study the naive continuum limit,
we define a continuum gauge potential at the boundaries through
\bea
U(x,k) & = & {\rm e}^{aA_k(x)} \,=\, \mathbf{1}_2 + aA_k(x) + {\rm O}(a^2) \,,
\eea
where $A_k=-ig_0A_k^3\sigma_3$. Here we use the continuum notation to
label the lattice points, $x_\mu=n_\mu a$.
Using the definition of lattice derivative
$\a(x+a {\hat k})=\a(x) + a\partial_k\a(x)$
and the properties $\a(x)^2=-\mathbf{1}_2$ and $\{A_k(x),\a(x)\}=0$
we get up to O($a^2$) corrections
\bea
 W_k(x) & = & 
 ia\tr\{\sigma_3\a(x)
 \underbrace{[\partial_k+2A_k(x)]}_{\displaystyle D_k}\a(x)\} \,.
\eea
Consistently with \eq{charge2}
the covariant derivative $D_k$ has a charge 2 in front of $A_k$. In summary,
in the naive continuum limit $W_k(x)$ corresponds to a covariant derivative
term for the Higgs field and is hence a vector in the representation of spin 1.

The photon mass can be extracted as follows: for each Higgs field $\Phi_i$
in the variational basis discussed above we construct
using \eq{photonfieldx} the corresponding photon field $W_k^{(i)}(n_0)$.
Then we build the matrix of connected correlation functions
\bea
 C^W_{ij}(t) & = &
 \frac{1}{3}\sum_k\left\{
 \left\langle \frac{a}{T}\sum_{n_0}W_k^{(i)}(n_0+t/a)W_k^{(j)}(n_0) \right\rangle
 \right. \nonumber \\
 && \left.
 -\left\langle \frac{a}{T}\sum_{n_0}W_k^{(i)}(n_0) \right\rangle
  \left\langle \frac{a}{T}\sum_{n_0}W_k^{(j)}(n_0) \right\rangle
 \right\} \,.
\eea
The expectation value of $W_k^{(i)}(n_0)$ is zero on average, nevertheless it is
subtracted to possibly reduce the statistical noise. The matrix $C^W_{ij}(t)$
is treated as its counterpart for the Higgs to finally get the masses.

%%%%%%%%%%%%%%%%%%%%%%%%%%%%%%%%%%%%%%%%%%%%%%%%%
\subsection{The static quark potential \label{ss_staticpot}}
%%%%%%%%%%%%%%%%%%%%%%%%%%%%%%%%%%%%%%%%%%%%%%%%%

We measure the potential between a static quark and a static antiquark placed
in the four-dimensional slices as we vary the location
$n_5=0,1,\ldots,\lfloor N_5/2\rfloor$ of the slice. The potential is extracted
from the expectation values of the traces of rectangular Wilson loops $W(t,r)$
of size $r$ in three-dimensional space and $t$ in time direction. The Wilson
loops are averaged over the three spatial directions and smeared gauge links,
constructed like it is explained in \sect{ss_higgsop}, are used for the
spatial Wilson lines. The potential $V(r)$ is then defined as the plateau value
at large time $t$ of the effective masses
\bea
 aV_{\rm eff}(r,t+a/2) & = & \ln\left(\frac{\tr\{W(r,t)\}}{\tr\{W(r,t+a)\}}\right) \,.
\eea
In the boundary slices of the orbifold the gauge links belong to gauge group $U(1)$,
whereas in the bulk slices they are $SU(2)$ gauge links. Since the links on the
boundary are ``colder'' (i.e. the boundary four-dimensional plaquettes have a larger
value) \cite{Knechtli:2005dw}, 
the potential turns out to be more precise on the boundary. Also, as the
value of $N_5$ is increased, the statistical errors on the potential also increase.
This calls for improving the extraction of the potential and there are methods
to do this.

%% file: sect4.tex
%%%%%%%%%%%%%%%%%%%%%%%%%%%%%
\section{Numerical results \label{s_numeric}}
%%%%%%%%%%%%%%%%%%%%%%%%%%%%%

We present detailed simulation results of the $SU(2)$ gauge theory on the
orbifold. The main part of the simulations is a scan in $\beta$ at fixed
$N_5=4$. The three-dimensional space size is $L/a=8$ and the temporal size
is $T/a=96$. The large value of the latter was necessary in order to extract
reliably the Higgs mass, which turns out to be very close to the 1-loop
perturbative value. To check for finite $L$ effects we also performed for
some $\beta$ values simulations at $L/a=12$. The algorithm is based on
$SU(2)$ heatbath and overrelaxation updates in the bulk and U(1) heatbath
and overrelaxation updates on the boundaries. The statistics of the
simulations at $L/a=8$ varies between 90'000 and 260'000 measurements
of the observables, at $L/a=12$ it is of 32'000 measurements.
Each measurement is separated by one heatbath and $L/2a$ overrelaxation
sweeps (i.e. updates of all the gauge links). 

We have also run some simulations to measure the static potential in the
four-dimensional slices along the extra dimension. For these runs the
orbifold geometry was chosen to be $T/a=32$ and $L/a=16$ and we compare
the potentials with $N_5=4$ and $N_5=6$ at $\beta=1.609$. This $\beta$
value was chosen, since the mass spectrum there resembles more what we
expect in a compactification scenario.

%%%%%%%%%%%%%%%%%%%%%%%%%%%%%%%%
\subsection{Phase transition \label{ss_pt}}
%%%%%%%%%%%%%%%%%%%%%%%%%%%%%%%%

%
%%%%%%%%%%%%%%%%%%%%%%%%%%%%%%%%%%%%%%%%%%%%%%%%%%%%%%%%%%%%%%%%%%%%%%%%%%%%%
\begin{figure}[!t]
\begin{minipage}{8cm}
\centerline{\epsfig{file=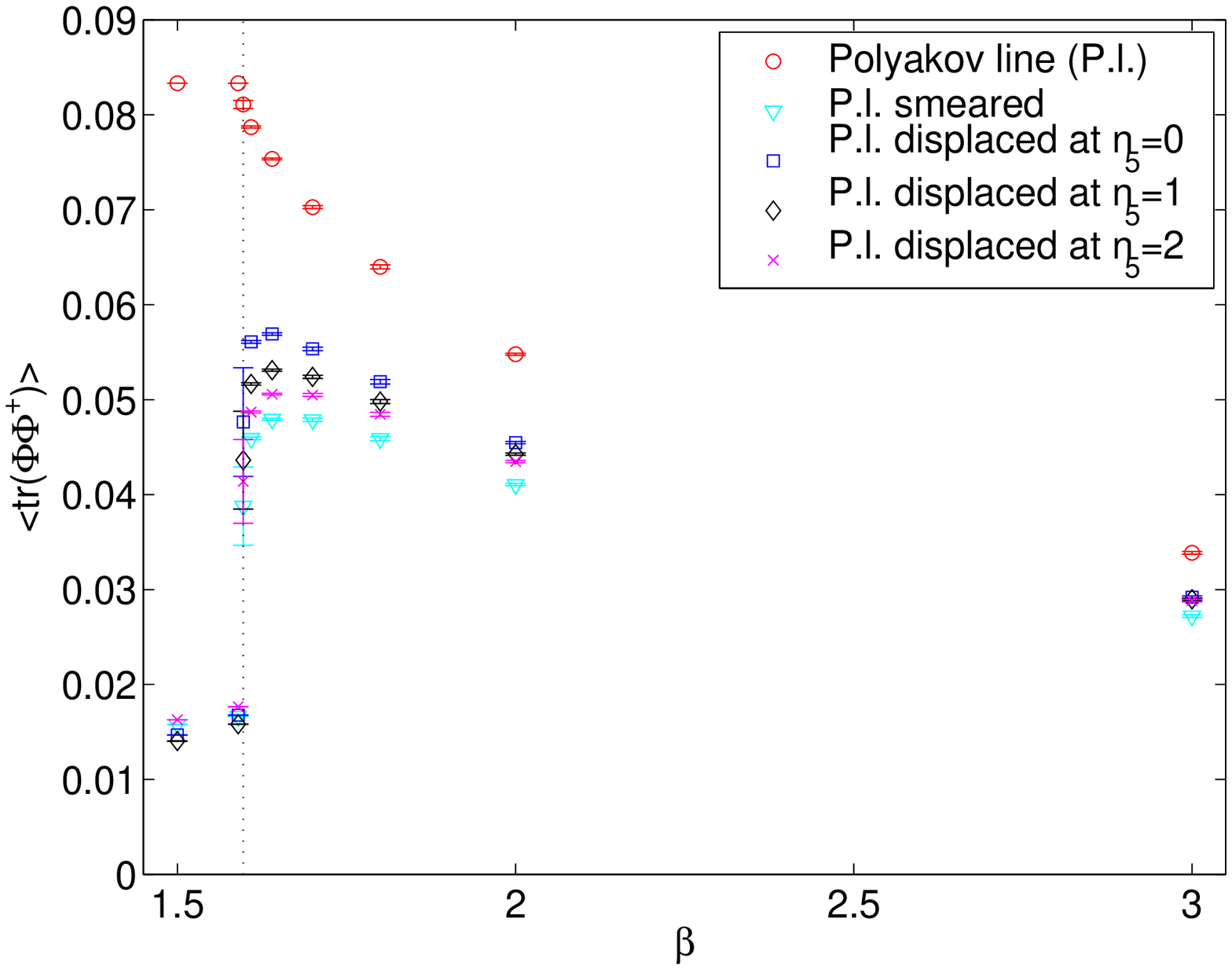,width=8cm}}
\end{minipage}
\begin{minipage}{8cm}
\centerline{\epsfig{file=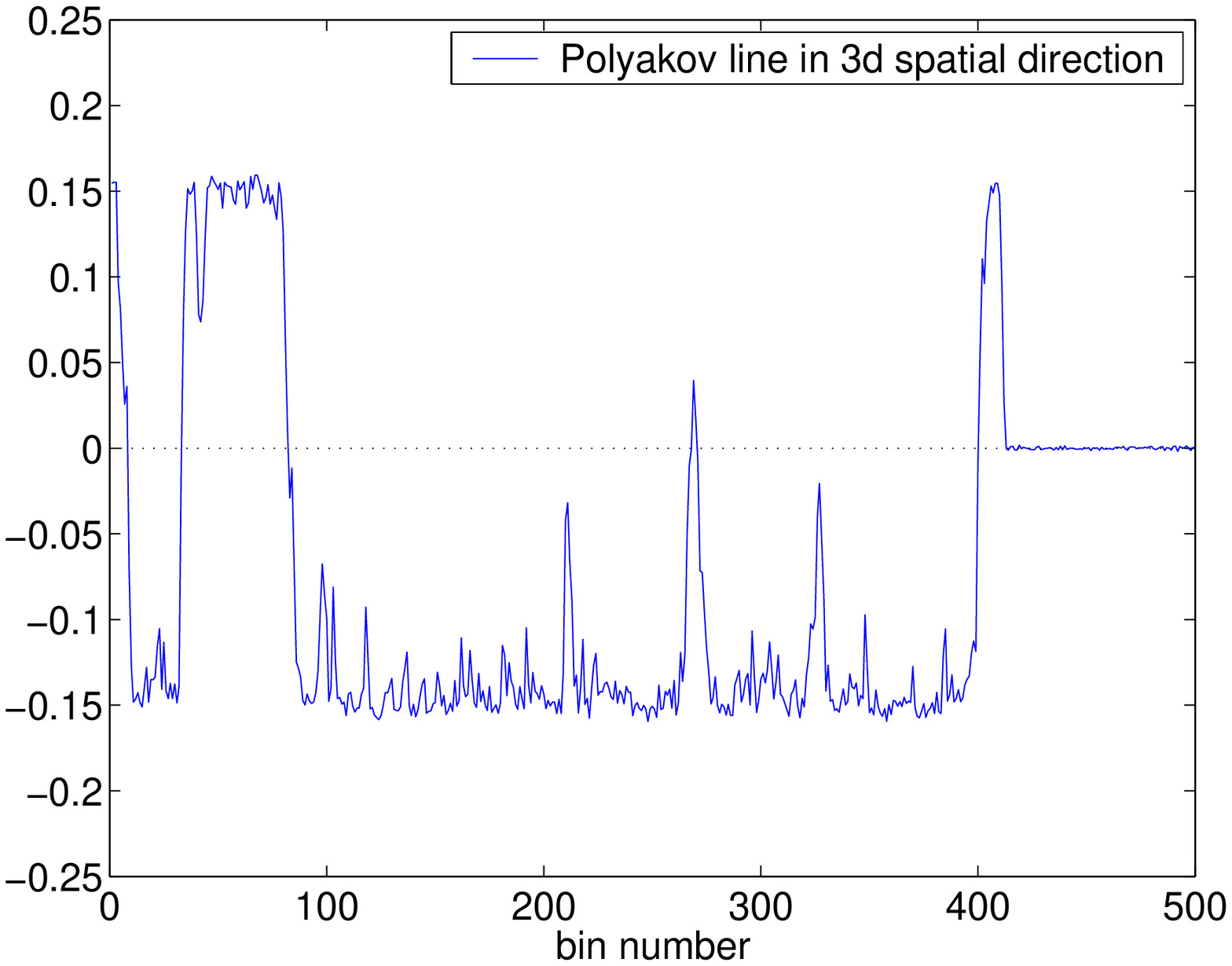,width=8cm}}
\end{minipage}
\caption{The phase transition on the orbifold at $N_5=4$. The plot
on the left hand side shows the behaviour of vacuum expectations values 
$\langle\tr\{\Phi^\dagger\Phi\}\rangle$ for different Higgs fields.
The plot on the right hand side shows a metastability in the Polyakov line
in one of the three-dimensional directions 
right at the phase transition $\beta_c=1.5975$.
\label{f_pt}}
\end{figure}
%%%%%%%%%%%%%%%%%%%%%%%%%%%%%%%%%%%%%%%%%%%%%%%%%%%%%%%%%%%%%%%%%%%%%%%%%%%%%
%

In infinite volume or with periodic boundary conditions on a finite torus
five-dimensional $SU(N)$ gauge theories have
at least two phases \cite{Creutz:1979dw,Beard:1997ic,Ejiri:2000fc},
a confinement massive phase at small values of $\beta$ and a deconfinement or
Coulomb massless phase at large values of $\beta$.
For $SU(2)$ the phase transition is located at $\beta_c=1.64$
\cite{Creutz:1979dw,Ejiri:2000fc}. With orbifold boundary conditions
this phase transition persists but the critical value $\beta_c$ depends
on $N_5$ and on $L/a$. It is signalled by a jump in the
expectation values of plaquettes, see \cite{Knechtli:2005dw}.
In \fig{f_pt} the plot on the left hand side shows what happens on the
orbifold with $N_5=4$ and $L/a=8$ with
vacuum expectation values $\tr\{\Phi\Phi^\dagger\}$, where for $\Phi$
we take some of the Higgs fields as explained in the legend and in
\sect{s_lat}. There is a clear discontinuity\footnote{
The expectation value of the basic Higgs field \eq{higgsfield} is actually
constant for $\beta<\beta_c$.} in the vacuum expectation
values at $\beta_c=1.5975$. 

There is a strong indication that the phase transition is of first order. 
The plot on the right hand side of \fig{f_pt}
shows the history\footnote{
The measurements for each simulations are blocked in 500 bins for
the error analysis.}
of the expectation value of the Polyakov line in one of the three-dimensional
space directions evaluated at the boundary $n_5=0$ of the orbifold.
As it was shown in early Monte Carlo study of $SU(2)$ gauge theory at
finite temperature \cite{McLerran:1980pk,Kuti:1980gh} the expectation
value of the Polyakov line is zero in the confined phase and it becomes
nonzero in the deconfinement phase, thereby breaking spontaneously a
$\mathbb{Z}_2$ symmetry which changed the sign of the Polyakov line.
In finite volume
one actually observes in the Monte Carlo history jumps between the
$\mathbb{Z}_2$ states. Precisely both of these behaviours can be seen on
the right hand side of \fig{f_pt}. Until about bin number 400 the system
was in the deconfined phase and then it changes into the confined phase.

The location of the phase transition depends not only on $N_5$ but also
on the ratio $L/a/N_5$. In principle we would like to be in a situation
where $N_5\ll L/a$ in order to have a compact extra dimension. But the
meaning of compactification will have to be qualified by looking at the
results for the particle spectrum.

%%%%%%%%%%%%%%%%%%%%%%%%%%%%%%%%
\subsection{Higgs and photon spectra \label{ss_spec}}
%%%%%%%%%%%%%%%%%%%%%%%%%%%%%%%%

%
%%%%%%%%%%%%%%%%%%%%%%%%%%%%%%%%%%%%%%%%%%%%%%%%%%%%%%%%%%%%%%%%%%%%%%%%%%%%%
\begin{figure}[!t]
\centerline{\epsfig{file=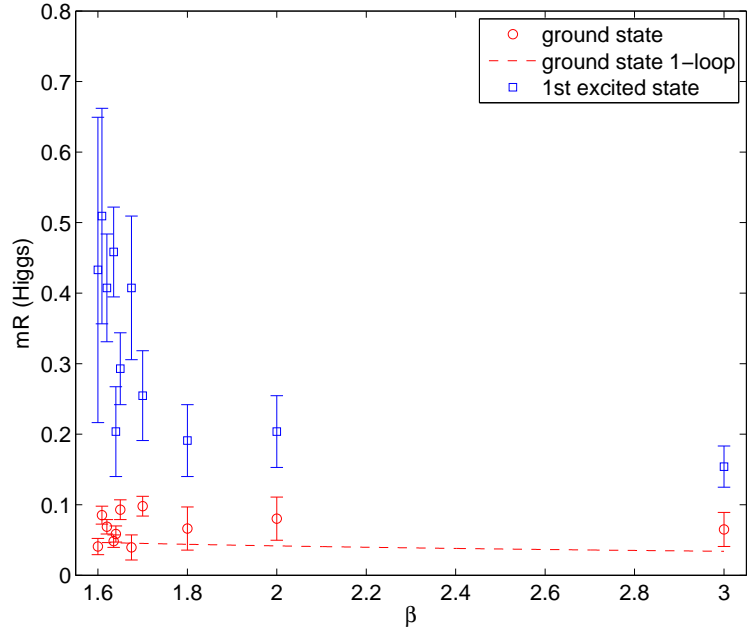,width=11.5cm}}
\caption{The ground and first excited state of the Higgs.
Scan in $\beta$ at fixed $N_5=4$. Masses are in units of
$1/R$. The dashed line represents the 1-loop result for the
ground state. \label{f_hispec}}
\end{figure}
\begin{figure}[!t]
\centerline{\epsfig{file=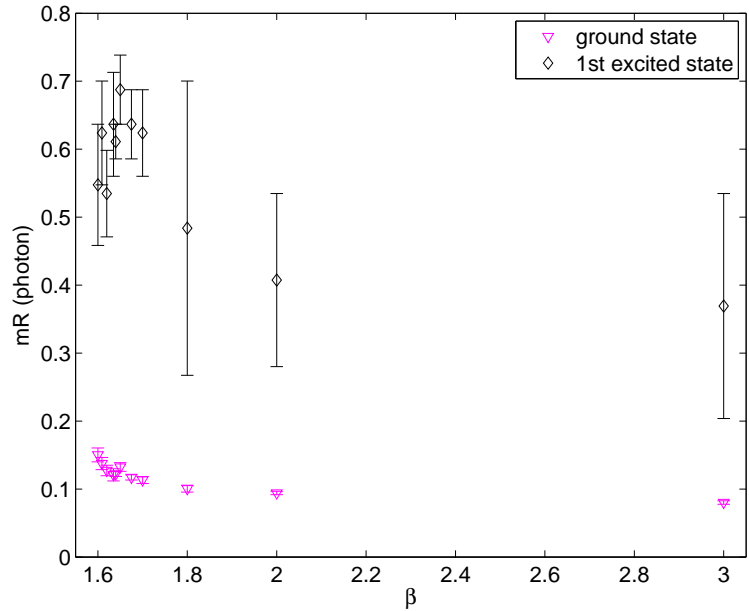,width=11.5cm}}
\caption{The ground and first excited state of the photon.
Scan in $\beta$ at fixed $N_5=4$. Masses are in units of
$1/R$. \label{f_phspec}}
\end{figure}
%%%%%%%%%%%%%%%%%%%%%%%%%%%%%%%%%%%%%%%%%%%%%%%%%%%%%%%%%%%%%%%%%%%%%%%%%%%%%
%

In \fig{f_hispec} and \fig{f_phspec}
the Higgs and photon masses for the ground state and the
first excited state are shown as a function of $\beta$ in units of $1/R$
for the $N_5=4$, $L/a=8$ and $T/a=96$ orbifold geometry. It was not possible
to extract these masses for $\beta<\beta_c=1.5975$. In the confined phase the
signal for the effective masses disappears immediately in noise. Right above
the phase transition the signal for the particle masses is there.

The first observation, looking at \fig{f_hispec},
concerns the Higgs ground state mass $m_h$.
For all $\beta>\beta_c$ the Higgs mass is
consistent with its value computed in 1-loop perturbation theory:
for general gauge group $G=SU(N)$ this is 
\bea
m_h R & = & \frac{c}{\sqrt{N_5 \b}} \,,\label{lattHiggs}
\eea
where $c=3/(4\pi^2) \sqrt{N\zeta(3)C_2(G)/2}$ and $C_2(G)=N$
($c=0.1178$ for $SU(2)$).
Here we have taken the continuum result in \cite{vonGersdorff:2002as}
and replaced the dimensionful coupling $g_5$ by its lattice definition
\eq{betalat}.

The second observation, looking at \fig{f_phspec},
concerns the photon ground state mass $m_\gamma$.
Contrary to the 1-loop prediction
\cite{Kubo:2001zc} the photon mass is non-zero for all $\beta>\beta_c$.
The photon mass even increases as the phase transition is approached.
This means that there is spontaneous symmetry breaking in the
pure gauge theory. This is, to our best knowledge, the first non-perturbative
evidence for the Higgs mechanism\footnote{
Spontaneous symmetry breaking in this context goes back to works
by Hosotani \cite{Hosotani:1983xw,Hosotani:1989bm}.}
originating from an extra dimension.

The third observation, looking at both \fig{f_hispec} and \fig{f_phspec},
concerns the excited state masses for the Higgs $m_h^\star$ and the photon
$m_\gamma^*$. In perturbation theory, the first excited (Kaluza--Klein)
states are expected to appear split from the ground states by
$(\Delta m)R=1$, the second with a mass splitting twice that, and so forth.
In no range of $\beta$ we see excited states at about 1 in units of $1/R$.
Close to the phase transition the excited states are separated from
the ground state, especially in the case of the photon.
Instead for larger $\beta$ they get closer in mass to the ground states.
This is an indication that at fixed $N_5=4$ the system behaves more
like a compact system close to the phase transition rather than for large $\beta$.

\begin{table}[!t] 
 \centering
  \begin{tabular}{lllll}
   \hline\\
   \multicolumn{1}{c}{$\beta$}  & 
   \multicolumn{1}{c}{$L/a$} & 
   \multicolumn{1}{c}{$m_h R$} & 
   \multicolumn{1}{c}{$m_\gamma R$} & 
   \multicolumn{1}{c}{$m_\gamma^* R$} \\[1.0ex]
   \hline\\
1.609 & 8  & 0.085(13) & 0.138(9)  & 0.62(8) \\
1.609 & 12 & 0.075(10) & 0.202(20) & 0.48(8) \\[0.5ex]

1.65  & 8  & 0.093(14) & 0.132(6)  & 0.69(5) \\
1.65  & 12 & 0.120(14) & 0.120(13) & 0.61(5) \\[0.5ex]

1.80  & 8  & 0.052(24) & 0.079(4)  & 0.38(17)\\
1.80  & 12 & 0.048(13) & 0.097(5)  & 0.70(6) \\[1.0ex]

   \hline
  \end{tabular} 
 \caption{Finite volume study of the spectrum at $N_5=4$.
 The excited state of the Higgs could not be determined at $L/a=12$.}
 \label{t_finvol}
\end{table}

At this point one might worry about finite $L$ effects in the particle
masses. Especially so for the photon mass, since its non-zero value contradicts
perturbation theory. In \tab{t_finvol} we offer a comparison at several
$\beta$ values of the particle masses between $L/a=8$ and $L/a=12$. One can
see that finite $L$ effects are small and in most cases not significant.
For sure there is no variation that could be explained with a behaviour
$m\propto1/L$ which is characteristic of finite volume effects.

%%%%%%%%%%%%%%%%%%%%%%%%%%%%%%%%
\subsection{Static potential in the four-dimensional slices \label{ss_potres}}
%%%%%%%%%%%%%%%%%%%%%%%%%%%%%%%%

%
%%%%%%%%%%%%%%%%%%%%%%%%%%%%%%%%%%%%%%%%%%%%%%%%%%%%%%%%%%%%%%%%%%%%%%%%%%%%%
\begin{figure}[!t]
\centerline{\epsfig{file=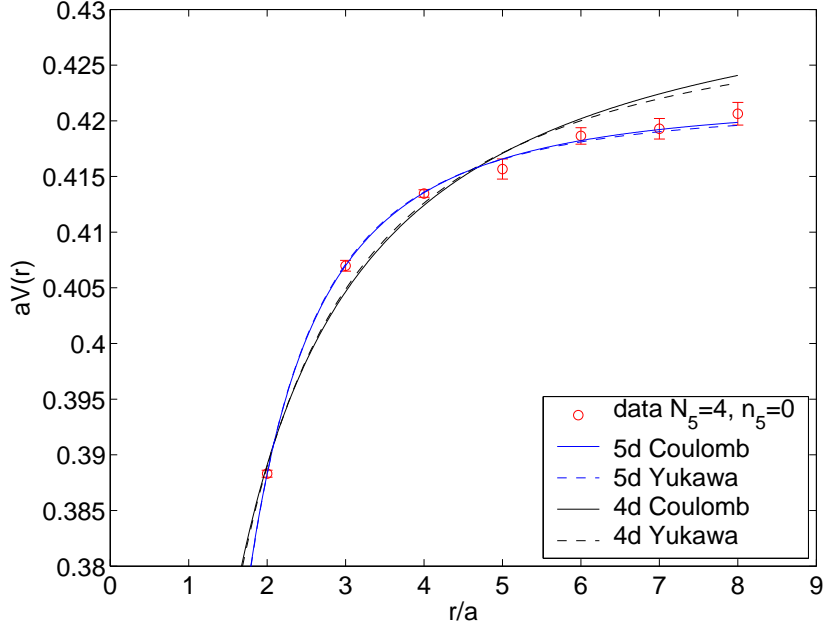,width=11cm}}
\caption{The static potential at $\beta=1.609$ and $N_5=4$
between static charges on the boundary.
The five-dimensional fits are clearly favoured. 
\label{f_potbN54}}
\end{figure}
%%%%%%%%%%%%%%%%%%%%%%%%%%%%%%%%%%%%%%%%%%%%%%%%%%%%%%%%%%%%%%%%%%%%%%%%%%%%%
%

In this section we present simulation data for the static potential in
the boundary slice, see \sect{ss_staticpot}, together with the results of
various fits. Since we know from the results in \sect{ss_spec} that the
gauge boson associated with the $U(1)$ gauge symmetry on the boundary
is massive, the physically motivated fits are Yukawa potentials
\bea
 aV(r) & = & -c_1\exp(-m_\gamma r)/(r/a)+c_0 \qquad \mbox{in four dimensions} \,,
 \label{Y4d} \\
 aV(r) & = & -d_1K_1(m_\gamma r)/(r/a)+d_0 \qquad \mbox{in five dimensions} \,,
 \label{Y5d}
\eea
where $c_0,c_1$ and $d_0,d_1$ are the fit parameters and the photon mass
$m_\gamma$ is the one measured in the simulations. In Appendix \ref{s_appb}
we derive the five-dimensional form of Yukawa potentials. It is interesting
to compare the Yukawa fits with Coulomb fits
\bea
 aV(r) & = & -f_1/(r/a)+f_0 \qquad \mbox{in four dimensions} \,,\label{C4d} \\
 aV(r) & = & -e_1/(r/a)^2+e_0 \qquad \mbox{in five dimensions} \,,\label{C5d}
\eea
where $f_0,f_1$ and $e_0,e_1$ are the fits parameters.
\begin{table}[!t] 
 \centering
  \begin{tabular}{lllll}
   \hline\\
   \multicolumn{1}{c}{$N_5$}  & 
   \multicolumn{1}{c}{5d Coulomb} & 
   \multicolumn{1}{c}{5d Yukawa} & 
   \multicolumn{1}{c}{4d Coulomb} & 
   \multicolumn{1}{c}{4d Yukawa} \\[1.0ex]
   \hline\\
4 & 0.4 & 0.6 & 14 & 10 \\[0.5ex]
6 & 1.1 & 1.9 & 4.2 & 1.7 \\[1.0ex]

   \hline
  \end{tabular} 
 \caption{The values of $\chi^2$/(degrees of freedom) for the various
 fits to the static potential at the boundary at $\beta=1.609$. A
 comparison between $N_5=4$ and $N_5=6$ is made.
 \label{t_potfits}}
\end{table}

In \fig{f_potbN54} we present results from a simulation of the orbifold
geometry $T/a=32$, $L/a=16$ and $N_5=4$. The statistics is of 20'000
measurements of the potential. The photon mass is
$am_\gamma=0.108(7)$ (for the fits we neglect its error), measured in the
simulations described in \sect{ss_spec}. In the fits only the points
$r/a=2,\ldots,8$ are included.
The $\chi^2$ per degree of freedom are listed in \tab{t_potfits}.
The four-dimensional fits are excluded whereas the five-dimensional ones
are very good.

%
%%%%%%%%%%%%%%%%%%%%%%%%%%%%%%%%%%%%%%%%%%%%%%%%%%%%%%%%%%%%%%%%%%%%%%%%%%%%%
\begin{figure}[!t]
\centerline{\epsfig{file=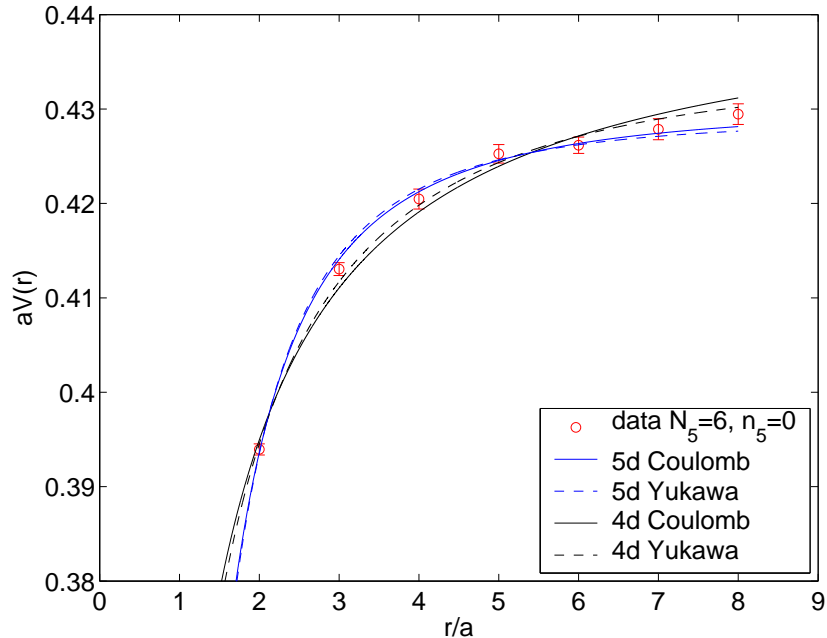,width=11cm}}
\caption{The static potential at $\beta=1.609$ and $N_5=6$
between static charges on the boundary.
Besides the five-dimensional fits also a four-dimensional
Yukawa fit can describe the data.
\label{f_potbN56}}
\end{figure}
%%%%%%%%%%%%%%%%%%%%%%%%%%%%%%%%%%%%%%%%%%%%%%%%%%%%%%%%%%%%%%%%%%%%%%%%%%%%%
%

The data at $N_5=4$ might suffer from the fact that the ratio between
the cut-off and the compactification scale is only $R/a=N_5/\pi\simeq1.3$.
In \fig{f_potbN56} we present results from a simulation of the orbifold
geometry $T/a=32$, $L/a=16$ and $N_5=6$. The statistics is of 9'000
measurements of the potential. The photon mass is
$am_\gamma=0.173(18)$, taken from a simulation at the same $\beta$ and
$N_5$ values but with $T/a=96$ and $L/a=12$.
In \tab{t_potfits} we can compare the $\chi^2$ per degree of freedom.
We see that at $N_5=6$ together with the five-dimensional fits also the
four-dimensional Yukawa form is a good fit. The four-dimensional Coulomb
fit is instead excluded.
It should be noted that the errors on the potential data are comparable
at both values of $N_5$.

We also measured the static potential in the middle slice at $n_5=2$.
The results of the fits confirm the conclusions from the fits in the
boundary slice at $N_5=4$, that is the four-dimensional forms are
disfavoured. The $\chi^2$ values are now about 3 for the
four-dimensional forms and 0.3 for the five-dimensional forms.
At $N_5=6$ the statistical errors are too large and all the fits
are equally good. As was discussed in \sect{ss_staticpot} we have to improve
the measurements of the static potential, especially for the middle
slice. It is important to have precise data for the latter, since any
difference with respect to the potential on the boundary might be
a hint of localization effects.

%%%%%%%%%%%%%%%%%%%%%%%%%%%%%%%%
\subsection{Discussion}
%%%%%%%%%%%%%%%%%%%%%%%%%%%%%%%%

A certainly surprising feature of these models emerges when we look at the mass spectrum itself,
measured in $R$-units. We have seen in \fig{f_hispec} that
when the compactification scale is lower than the cut-off scale by a fixed gap (i.e. for $N_5$ fixed),
the Higgs ground state mass is consistent with its 1-loop perturbative value
not only at large values of $\beta$ but also in the definitely
non-perturbative region when $\beta$ is close to its critical value at
the phase transition. Moreover also the photon mass has a mild dependence
on $\beta$, see \fig{f_phspec}.
Since we do not have at present any powerful analytical tools to explain
these observations, we will 
keep our discussion at a qualitative level. 

We would like to argue that 
one way to interpret this peculiar phenomenon is by a very mild
cut-off dependence of the five-dimensional bare coupling $g_5^2$ in the non-perturbative regime\footnote{Note that this is not in contradiction with the  
perturbative, 1-loop
expectation which wants the five-dimensional bare coupling to have a  
mild cut-off dependence for energy scales $E \simeq 1/R$ where the theory
is truly five-dimensional (see eq. (3.13) of \cite{Dienes:1998vg}).}.
The generic situation observed in our simulations, as far as the spectrum
is concerned, can be translated in
an effective field theory language by saying that
the theory possesses a region in its parameter space such that any  
operator of dimension $5+p$ appearing in the effective action and 
effective operators\footnote{
This is Symanzik's analysis of cut-off effects.}
scales as
\bea
\frac{{\cal Z}(g_0, g_4)}{\Lambda^p}{\cal O}^{(5+p)} \,, \label{effop}
\eea
with ${\cal Z}(g_0, g_4)$ a very slowly varying function of the
dimensionless bare couplings $g_0=g_5\sqrt{\Lambda}$ and $g_4=g_5/\sqrt 
{\pi R}$, at least as long as
the ratio $\Lambda/(1/R)$ is kept fixed. In the orbifold theory, taking  
the Higgs mass as an example,
we recall that a boundary counterterm is non-perturbatively excluded  
and thus all
possible corrections come from either pure bulk effects or bulk  
effects with boundary insertions,
both of which descend directly from the circle.

In an attempt to predict the explicit cut-off dependence in \eq{effop},
we note that naive dimensional analysis tells us
that as $\b$ decreases with $g_5$ fixed, the cut-off increases,
see \eq{betalat}.
The compactification scale is $1/R$ and
a wide separation from the cut-off scale requires $\Lambda >> 1/R$.
Increasing $\b$ while keeping the gap between the compactification and
cut-off scales fixed, would require decreasing $\Lambda$ (at fixed $g_5$)
and therefore an increase of $R$, which drives the fifth dimension to its  
decompactification limit.
A general lesson then is that for fixed $N_5$, moving towards
the large $\b$ regime is expected to enhance the cut-off
effects (appearing as $E/\Lambda$ at low energies $E$ in the sense
of an effective action) and decompactify the theory,
whereas moving in the opposite direction, i.e. towards smaller $\b$,
is expected to suppress the cut-off effects and drive the system into  
a compactified but
non-perturbative regime.
Eventually the phase transition is reached at the critical value of $\b=\b_c$, where the cut-off  
reaches its maximal value.
There is a possible caveat though in this argument: we have implicitly
assumed that it makes sense to vary the cut-off while keeping the  
dimensionful
coupling $g_5$ fixed for all values $\beta_c < \b < \infty$.

If we look at the spectrum of the excited states of the Higgs and the
photon in \fig{f_hispec} and \fig{f_phspec}, we see that as $\beta$
grows their masses decrease, thus deviating more and more from their
expected perturbative value as Kaluza--Klein states of mass $\simeq1/R$.
This fact could be explained with our argument that cut-off effects
increase at large $\beta$. We cannot at the moment explain why these
cut-off effects would affect the excited states but not the ground states.

Next, let us see what happens when we start changing $N_5$ while keeping
$\b$ fixed. The cut-off $\Lambda$ depends only on $\beta$ and is also fixed.
Using \eq{Rlat} we obtain that
\be
\frac{\d N_5}{N_5} = \frac{\d R}{R}
\ee
which implies that increasing $N_5$ amounts to increasing also $R$.
Thus, in a compactified scenario we expect to be able to approach an effective dimensional  
reduction by decreasing $N_5$. 
It is interesting to note that according to our potential data we observe
an opposite effect: the fit to a four dimensional Yukawa law becomes
much better when we increase $N_5$ from 4 to 6 at fixed $\beta$, namely when 
we decompactify the extra dimension. 
Therefore, to the extent that this result can be considered as a firm physical 
property of the system in this region of the parameter space, 
we are lead to the conclusion that the fact that we start observing an effective dimensional
reduction at $N_5=6$ is more likely to be a consequence of a localization mechanism
rather than an effect of compactification.
It is possible that the region that corresponds to 
dimensional reduction from compactification is located
at much smaller $N_5$ which would however require an anisotropic lattice
to be probed.

Finally, regarding the applicability of the vicinity of the phase transition 
in model building, we would like to point out
that even though it clearly corresponds to a strong coupling regime from the point of view
of five dimensions, viewed from the point of view of the four-dimensional effective theory
with an effective coupling defined as in \eq{b4},
it could correspond to a weakly coupled regime, if for example
large renormalization factors change \eq{b4}.

%% file: sect5.tex
%%%%%%%%%%%%%%%%%%%%%%%%%%%%%%%%%%
\section{Conclusions \label{s_concl}}
%%%%%%%%%%%%%%%%%%%%%%%%%%%%%%%%%%

The first results for the spectrum of the orbifold theory simulated
on a lattice with an extra dimension of the size of $N_5=4$ lattice spacings,
show that there is spontaneous symmetry breaking, which manifests in a massive
gauge boson associated with the $U(1)$ boundary gauge group.
This result was unexpected from computations at 1-loop in perturbation
theory, where the photon remains massless. Moreover, the Higgs mass is
also measured in a large range of gauge coupling and it is always close
to its perturbative value. The excited state masses for the Higgs and
the photon are not at the scale of the inverse compactification radius,
where they are expected to be in a scenario of dimensional reduction
like in finite temperature field theory. But close to the orbifold phase
transition the mass splitting between the ground states and the first
excited states increases. Data for the static potential strongly suggest
five-dimensional potential forms, both when the static potential is
measured with the charges on the boundary slice or in the middle slice
along the extra dimension. We have also presented potential data at $N_5=6$.
They indicate that a four-dimensional Yukawa form, with the vector boson
mass equal to the measured photon mass, is a good fit to the potential
on the boundary together with the five-dimensional forms.

The results we have so far give a consistent picture and they motivate
for further work to explore the phase diagram of the orbifold theory.
Despite its non-renormalizability, the theory makes finite predictions
for the mass spectrum and we would like to understand better how to
change the lattice parameters to study the scaling properties.
This also requires technical improvements, for the basis of Higgs
operators and for the static potential.

%% file: appa.tex
%%%%%%%%%%%%%%%%%%%%%%%%%%%%%
\section{Higgs potential from extra dimension(s) \label{s_appa}}
%%%%%%%%%%%%%%%%%%%%%%%%%%%%%

In this Appendix we review the calculation of 1-loop potentials in orbifold gauge theories.
This material is already well known in the hep-ph community but since it is perhaps
less familiar in the lattice community, for completeness 
we reproduce here in detail the perturbative arguments
for the existence or not of spontaneous symmetry breaking, reproducing 
essentially some of the the results of \cite{Kubo:2001zc}. 

Consider a (massive) free field theory for a one--component real scalar field
in D-dimensions with action $S$.
After a Euclidean rotation, the path intergral that defines the vacuum energy
$\Gamma$ is \cite{Antoniadis:2001cv}:
\bea
{\rm e}^{-\Gamma} = 
\int [D\phi]\; e^{-S_E} \sim \frac{1}{\sqrt{\det{[\Box + M^2] }}}
\,,\quad \Box=-\partial_\mu\partial_\mu\,.
\eea
The mass (M) dependence of the above can be extracted by using the identity
\bea
\log \left( \det A \right) = - \int _\e ^\infty \frac{{\rm d}t}{t} 
\tr \left({\rm e}^{-tA} \right).
\eea
We obtain
\bea
\G = - \log \left[ \frac{1}{\sqrt{\det{[\Box + M^2] }}}\right] = -\frac{1}{2}\frac{\pi^{D/2}}{(2\pi)^D}\int_0^\infty
\frac{{\rm d}t}{t^{\frac{D+2}{2}}}{\rm e}^{-tM^2}
\ea
For a fermionic free field we have
\bea
{\rm e}^{-\Gamma} = \det(D) \,,\quad D=\gamma_\mu\partial_\mu+M \,.
\eea
The nonzero eigenvalues of the Dirac operator come in complex conjugate pairs
$\pm i\alpha+M$, hence
\bea
\det(D) = \sqrt{\det(DD^\dagger)} = \sqrt{\det[\Box+M^2]} \,.
\eea
We can therefore summarize the contributions to the
effective action, setting $D=4$, as
\bea
\Gamma
&=& - \frac{1}{32\pi^2} \sum_I (-)^{F_I} \int_0^\infty {\rm d}l \; l\;  
{\rm e}^{-\frac{M_I^2}{l}}\,, 
\eea
where $F_I$ is 0 for bosonic and 1 for fermionic degrees of freedom
of mass $M_I$.

Let us apply the above for the case where there are $d$ extra 
compact toroidal directions.
For the four-dimensional theory, after dimensional reduction, i.e.
Kaluza--Klein decomposition, we obtain the mass formula
\bea
M^2_{I} = m^2_{I} + \sum_{i=1}^d \left(\frac{n_i+a_i^I}{R_i}\right)^2,
\label{massformula}
\eea
where
\begin{itemize}
\item
$R_i$ are the radii of the $d$ circles of the torus.
\item
$\{n_i\},\;i=1,\ldots,d$ are integers.
\item
$m^2_{I}$ is a $4+d$-dimensional mass which remains in the $R_i\longrightarrow \infty$
limit. This is zero for a pure gauge theory.
\item
The shifts $a_i^I$ originate from the possible failure for periodicity of
the $4+d$ dimensional field $\Phi_I$:
\bea
\Phi_I (x^\m, y^i + 2\pi k_i R_i) = e^{2 i \pi \sum_i k_i a_i^I} \Phi_I (x^\m, y^i), 
\eea
where $y_i$ are the coordinates of the circles. 
\end{itemize}
According to the above, for the $T^d$ torus we have the expression for the potential
\bea
V_{eff}^{T^d} = - \sum_I \sum_{\{n_i\}} 
(-)^{F_I} \frac{1}{32\pi^2} \int_0^\infty
{\rm d}l\, l\, e^{-\sum_i \frac{(n_i+a_i^I)^2}{R_i^2 l}}.
\eea
By commuting the integral with the sum over $n_i$ and then performing a Poisson
resummation using the formula
\bea
\sum _{\{n_i\}} e^{{-\pi\; {\bf n} ^T {\bf A} {\bf n} } + 2 i \pi {\bf b}^T {\bf n}}=
\frac{1}{\sqrt{\det{\bf A}}} \sum _{\{m_i\}}
e^{-\pi\; ({\bf m-b}) ^T {\bf A}^{-1} ({\bf m-b}) }
\eea
where ${\bf A}$ is a $d\times d$ (invertible) matrix and ${\bf n}$ and ${\bf m}$ 
are $d$ dimensional KK number vectors, we find 
\bea
V_{eff}^{T^d} = - \sum_I (-)^{F_I} 
\frac{\left(\prod_i R_i\right)}{32\pi^{\frac{4-d}{2}}}
\sum _{\{n_i\}} {\rm e}^{2 i \pi \sum_i n_i a_i^I} 
\int _0^\infty {\rm d}l\, l^{\frac{2+d}{2}}\, 
e^{-\pi^2l\sum_in_i^2R_i^2}.
\eea
The terms with $n_i=0$ give rise to a divergent contribution to the vacum energy.
They represent a contribution to the cosmological constant and can be neglected for the
present discussion.
For all other non-vanishing vectors $\vec n$
we perform the integral explicitly. This leads to the finite result 
\bea
V_{eff}^{T^d} = - \sum_I (-)^{F_I} 
\frac{\Gamma\left(\frac{4+d}{2}\right)}{32\pi^{\frac{12+d}{2}}}
\left(\prod_i R_i\right) \sum _{\vec n \ne 0} 
\frac{e^{2\pi i \sum_i n_i a_i^I}}{\left(\sum_i n_i^2 R_i^2 \right)^{\frac{4+d}{2}}}.
\eea
%
 
%
%%%%%%%%%%%%%%%%%%%%%%%%%%%%%%%%%%%%%%
\subsection{5D $SU(2)$ gauge theory on $S^1/Z_2$}
%%%%%%%%%%%%%%%%%%%%%%%%%%%%%%%%%%%%%%
%

For a 5D theory compactified on the 1 dimensional torus, the circle,
we have $d=1$. Then
\bea
V_{eff}^{S^1} = - \sum_I (-)^{F_I} \frac{\Gamma\left(\frac{5}{2}\right)}{32\pi^{\frac{13}{2}}}\, R\, 
\sum _{n \ne 0} 
\frac{e^{2\pi i n a^I}}{\left(n^2 R^2 \right)^{\frac{5}{2}}}
= - \frac{3}{64 \pi^6 R^4}\sum_I (-)^{F_I} 
\left[ \sum_{n=1}^\infty \frac{\cos{(2\pi n a^I)}}{n^5}\right].
\eea
For a pure gauge theory we have in addition that $F_I=0$.
We then finally obtain for the vacuum energy
\bea
V_{eff}^{S^1} = - \frac{3}{64 \pi^6 R^4} \sum_A \sum _{n=1}^\infty
\frac{\cos{(2\pi n a^A)}}{n^5}\label{vacenergy}
\eea
where the index $I$ has been changed to $A$, indicating the adjoint
representation of some gauge group.
This potential may have a minimum that leads to
an extra dimensional version of the Higgs mechanism.

There are different cases where a failure of periodicity can happen and a 
shift in the KK formula is generated. One particularly interesting example
is the presence of a Wilson line in an extra dimensional gauge theory
\bea
a^A = q^A\, g_5\, \int_{S^1} \frac{dy}{2\pi} \, A_5^A,
\eea
where $A_5^A$ is the internal component of the gauge field with gauge coupling $g_5$ 
and $q_A$ the charge of the $A$th field under the corresponding generator.
To be more concrete, let us look at an example.
Consider a 5D $SU(2)$ gauge theory compactified on the circle.
In order to compute the vacuum energy, according to eq.
(\ref{vacenergy}) the only quantity we need to compute 
are the $a^A$. To compute the $a^A$ we have to compute the eigenvalues
of the mass-squared operator \cite{Kubo:2001zc}
\bea
-D_M D_M = -D_\m  D_\m - D_5 D_5 \quad \mbox{(in Euclidean space)}
\eea
in the vacuum, with $D_M$ acting on fields $F=F^AT^A$ in the adjoint representation
as
\bea
D_MF & = & \partial_MF + [B_M,F]\,,\quad B_M=\langle A_M\rangle\,.
\eea
$B_M$ is a constant background field given by the vacuum expectation value (vev) of 
$A_M$.

After orbifolding, which breaks $SU(2)\longrightarrow U(1)$
the only fields that can take a vev are 
$A_5^1 (x^5)$ and $A_5^2 (x^5)$. The unbroken $U(1)$ invariance can be used
to rotate the vev's in such a way that we have
\bea
B_5^1=H\,,\quad B_5^2=B_5^3=0 \,,\quad B_5=-ig_5B_5^AT^A \,.
\eea
The non-trivial part comes from the $D_5 D_5$ part of the operator acting
on $x^5$ dependent parts of the gauge fields. Writing out this 
explicitly in the vacuum acting on an adjoint field $F$,
\bea
 D_5D_5F & = & \partial_5\partial_5F + 2[B_5,\partial_5F] + [B_5,[B_5,F]] \,,
\eea
gives the operator
\bea
\left( D_5 D_5 \right)_{AB} = 
\pmatrix {\partial_5 \partial_5 & 0 & 0 \cr
0 & \partial_5 \partial_5 - g_5^2 HH & -2g_5 H \partial_5\cr
0 & 2g_5 H \partial_5 & \partial_5 \partial_5 - g_5^2 HH}.
\eea
Since this operator does not mix different KK modes we can diagonalize
it separately for each level $n$. 
\begin{itemize}
\item
Eigenvalues of $A_\m^A$.
                                                                                        
The matrix elements of the $- D_5 D^5$
operator 
\bea
\langle f|-D_5D_5|g\rangle & = & 
\int_0^{2\pi R}{\rm d}x^5\,f^*(x^5)(-D_5D_5)g(x^5)
\eea
can be easily obtained by using the basis of orthonormal functions
\bea
\frac{1}{\sqrt{\pi R}} \cos {\frac{n}{R}x^5},\hskip .5 cm {\rm for}\; A,B=3
\eea
since $A_{\m}^3$ is even under the orbifold, and
\bea
\frac{1}{\sqrt{\pi R}} \sin {\frac{n}{R}x^5},\hskip .5 cm {\rm for}\; A,B=1,2
\eea
since $A_{\m}^{1,2}$ are odd under the orbifold.
The matrix we obtain for $n\ne 0$ is
\bea
A_\m^{A\, (n\ne 0)}\longrightarrow \pmatrix {\frac{n^2}{R^2} & 0 & 0 \cr
0 & \frac{n^2}{R^2} + \frac{\a^2}{R^2} & -2\frac{\a n}{R^2} \cr
0 & -2\frac{\a n}{R^2} & \frac{n^2}{R^2} + \frac{\a^2}{R^2}}\longrightarrow
\frac{n^2}{R^2}, \frac{(|n|+\a)^2}{R^2}, \frac{(|n|-\a)^2}{R^2} \,,
\label{shiftsamu}
\eea
where
\bea
\a = g_5\, H\, R
\eea
and on the right we have given the eigenvalues.
For $n=0$, the $A_\m^{1,2}$ do not have zero modes but
$A_\m^3 $ does have a zero mode whose eigenvalue is
\bea
A_\m^{3,(0)}\longrightarrow \frac{\a^2}{R^2}.
\eea
\item
Eigenvalues of ghosts.

We notice that the ghosts have the same $Z_2$ parity
assignement as the gauge fields $A_\m$ therefore 
the effect of the ghosts is to just reduce 
the degree of polarization as $4\rightarrow 2$.
\item
Eigenvalues of $A_5^A$. 

Since the $Z_2$ parities of the $A_5^A$ are the opposite from those
of $A_\m^A$, we will obtain 
\bea
A_\m^{1,2,(0)}\longrightarrow \frac{\a^2}{R^2}
\eea
and 
\bea
A_5^{A\, (n\ne 0)}\longrightarrow 
\frac{n^2}{R^2}, \frac{(|n|+\a)^2}{R^2}, \frac{(|n|-\a)^2}{R^2} \,.
\label{shiftsa5}
\eea
\end{itemize}
By comparing with \eq{massformula} we see that non-zero shifts $a^I$
are given by $\pm\alpha$ in \eq{shiftsamu} and \eq{shiftsa5}.
As we have argued, only the non-zero modes are relevant for the 
vacuum energy. We count $2+1$ (2 from physical degrees of polarization of $A_\mu$,
1 from $A_5$) eigenvalues $(n+\a)^2/R^2$ (with $a^I=\a$)
and an equal number of eigenvalues $(n-\a)^2/R^2$ (with $a^I=-\a$). 
Furthermore, as can be seen from eq. (\ref{vacenergy})
the $a^I=-\a$ contribution is the same as the $a^I=+\a$ contribution.
We then obtain for the vacum energy the final result
\bea
V_{eff}^{S^1}(SU(2)) = - \frac{3\cdot 3\cdot 2}{64 \pi^6 R^4} \sum _{n=1}^\infty
\frac{\cos{(2\pi n \a)}}{n^5} \,.\label{potSU2}
\eea
This potential could, in principle break the remaining $U(1)$ down to nothing on the branes.
Plotting the potential one can see that it has two degenerate minima at
$\a=0$ and $\a=1$, none of which breaks $U(1)$.

A similar computation gives for the $SU(3)\longrightarrow SU(2)\times U(1)$ 
model the potential
\bea
V_{eff}^{S^1}(SU(3)) = - \frac{3\cdot 3\cdot 2}{64 \pi^6 R^4} \sum _{n=1}^\infty
\frac{1}{n^5} \left[ \cos{(2\pi n \a)} + 2 \cos{(\pi n \a)} \right].
\eea
The latter can be seen by noticing that the eigenvalues 
for the adjoint of $SU(3)$ are
\bea
2\times \frac{n^2}{R^2},\hskip .5cm \frac{(n\pm \a)^2}{R^2}
\hskip .5cm 2\times \frac{(n\pm \frac{\a}{2})^2}{R^2}.
\eea
Again, the minimum is at $\a=0$ which does not break 
$SU(2)\times U(1)$.

In fact, in some cases a useful criterion to see if it is possible to
break the symmetry by the Hosotani mechanism is to use the statement
proven in \cite{vonGersdorff:2002as}:
The Hosotani mechanism does not reduce the rank of ${\cal H}$ if the symmetry
breaking global minimum is at $\a=1$.  To show this
statement, we can compute the Wilson line due to the vev $H=\frac{1}{g_5R}$
of a scalar along the $T^A$ direction:
\begin{equation} \langle W\rangle =
{\mathcal P}{\rm e}^{-ig_5\oint {\rm d}x^5\,H\,T^A} =
{\rm e}^{-2\pi i T^A} \,.\end{equation} 
It is straightforward to show that $\exp(-2\pi i T^{A})$ is a diagonal
matrix. In the case of $SU(2)$: $\exp(-i\pi\sigma_1)=-\mathbf{1}_2=\exp(-i\pi\sigma_2)$.
For $SU(N)$, the non-diagonal generators are obtained embedding $\sigma_{1,2}$. 
Thus, we always have that
\begin{equation}  [\langle W\rangle ,H_i] =0,\end{equation}
where $H_i$ are the generators corresponding to the Cartan subalgebra
of ${\cal G}$, i.e. that the Wilson loop commutes with at least those
generators and therefore it leaves at least a $U(1)_1\times \cdots
\times U(1)_{rank({\cal H})}$ unbroken. 
From this it is clear that in the $SU(2)$ model, $U(1)$ can not break further with
$\a=1$. For the $SU(3)$ model $\a =1$ could break $SU(2)\times U(1)$
down to $U(1)\times U(1)$ by the Hosotani mechanism. 
In both cases, it is not clear what $\a \ne 1$ would do.
 

%% file: appb.tex
%%%%%%%%%%%%%%%%%%%%%%%%%%%%%
\section{The Yukawa potential in 5D \label{s_appb}}
%%%%%%%%%%%%%%%%%%%%%%%%%%%%%

The potential is given by the integral
\be 
V(x) = \int \frac{d^4q}{(2\pi)^4}\frac{e^{iq \cdot x}}{q^2 + m^2}
\ee
where $|x|=r$, $q=|q| q'$, $|q'|=1$ and
\bea
&& q_1' = \cos{\phi_1}\nonumber\\
&& q_2' = \sin{\phi_1}\cos{\phi_2}\nonumber\\
&& q_3' = \sin{\phi_1}\sin{\phi_2}\cos{\theta}\nonumber\\
&& q_4' = \sin{\phi_1}\sin{\phi_2}\sin{\theta}
\eea 
where $0\le \phi_j \le \pi, 0\le \theta \le 2\pi$. Then the potential can be written as
\be
V(x) = \frac{1}{(2\pi)^4} \int_0^\infty d|q| |q|^3 \int_0^{2\pi} d\theta \int_0^\pi d\phi_1
\int_0^\pi d\phi_2 \sin^2\phi_1 \sin\phi_2 \frac{e^{i q\cdot x}}{q^2+m^2}.
\ee
Choosing coordinates such that $x=r\cdot e_1$ with $e_1$ a unit vector, we have that $q\cdot x = |q| r \cos\phi_1$ and then
\be
V(x) = \frac{1}{(2\pi)^4}\int_0^\infty d|q| \frac{|q|^3}{|q|^2+m^2}\int_0^{2\pi}d\theta
\int_0^\pi d\phi_2\sin\phi_2 \int_0^\pi d\phi_1\sin^2\phi_1e^{i|q|r\cos\phi_1}.
\ee
Performing the angular integrals we obtain
\be
V(x) = \frac{1}{(2\pi)^2}\frac{1}{r} \int_0^\infty d|q| \frac{|q|^2}{|q|^2+m^2}J_1(|q|r).
\ee
Finally, changing variables as $y=|q|r$ one has the left over radial integral
\be
V(x)=\frac{1}{(2\pi)^2r^2}\int_0^\infty dy \frac{y^2}{y^2+(mr)^2}J_1(y),
\ee
which can be done, yielding the result
\be
V(x) \sim \frac{1}{(2\pi)^2r^2} (mr) K_1(mr).
\ee
As $r\longrightarrow 0$, the Bessel function $(mr) K_1(mr)\longrightarrow 1$ so we have at short distances
\be
r\longrightarrow 0:\hskip 1cm V(x) = \frac{1}{(2\pi)^2r^2}.
\ee
For large $r$ on the other hand we have 
\be
r\longrightarrow \infty:\hskip 1cm V(x) = \sqrt{\frac{{\pi m}}{2}}\frac{1}{(2\pi)^2} \frac{e^{-mr}}{r^{3/2}}.
\ee

%% file: LatticeHiggs1.bbl
\begin{thebibliography}{10}

\bibitem{Coleman:1973jx}
S.R. Coleman and E. Weinberg,
\newblock Phys. Rev. D7 (1973) 1888.

\bibitem{Fairlie:1979at}
D.B. Fairlie,
\newblock Phys. Lett. B82 (1979) 97.

\bibitem{Fairlie:1979zy}
D.B. Fairlie,
\newblock J. Phys. G5 (1979) L55.

\bibitem{Manton:1979kb}
N.S. Manton,
\newblock Nucl. Phys. B158 (1979) 141.

\bibitem{Forgacs:1979zs}
P. Forgacs and N.S. Manton,
\newblock Commun. Math. Phys. 72 (1980) 15.

\bibitem{Hosotani:1983xw}
Y. Hosotani,
\newblock Phys. Lett. B126 (1983) 309.

\bibitem{Hosotani:1989bm}
Y. Hosotani,
\newblock Ann. Phys. 190 (1989) 233.

\bibitem{Antoniadis:2001cv}
I. Antoniadis, K. Benakli and M. Quiros,
\newblock New J. Phys. 3 (2001) 20, hep-th/0108005.

\bibitem{Dvali:2001qr}
G.R. Dvali, S. Randjbar-Daemi and R. Tabbash,
\newblock Phys. Rev. D65 (2002) 064021, hep-ph/0102307.

\bibitem{Arkani-Hamed:2001nc}
N. Arkani-Hamed, A.G. Cohen and H. Georgi,
\newblock Phys. Lett. B513 (2001) 232, hep-ph/0105239.

\bibitem{vonGersdorff:2002as}
G. von Gersdorff, N. Irges and M. Quiros,
\newblock Nucl. Phys. B635 (2002) 127, hep-th/0204223.

\bibitem{Cheng:2002iz}
H.C. Cheng, K.T. Matchev and M. Schmaltz,
\newblock Phys. Rev. D66 (2002) 036005, hep-ph/0204342.

\bibitem{Arkani-Hamed:2001mi}
N. Arkani-Hamed, L.J. Hall, Y. Nomura, D.R. Smith and N. Weiner,
\newblock Nucl. Phys. B605 (2001) 81, hep-ph/0102090.

\bibitem{Masiero:2001im}
A. Masiero, C.A. Scrucca, M. Serone and L. Silvestrini,
\newblock Phys. Rev. Lett. 87 (2001) 251601, hep-ph/0107201.

\bibitem{vonGersdorff:2002rg}
G. von Gersdorff, N. Irges and M. Quiros,
\newblock (2002), hep-ph/0206029.

\bibitem{vonGersdorff:2002us}
G. von Gersdorff, N. Irges and M. Quiros,
\newblock Phys. Lett. B551 (2003) 351, hep-ph/0210134.

\bibitem{Irges:2004gy}
N. Irges and F. Knechtli,
\newblock Nucl. Phys. B719 (2005) 121, hep-lat/0411018.

\bibitem{Martinelli:2005ix}
G. Martinelli, M. Salvatori, C.A. Scrucca and L. Silvestrini,
\newblock (2005), hep-ph/0503179.

\bibitem{Lim:2006bx}
C.S. Lim, N. Maru and K. Hasegawa,
\newblock (2006), hep-th/0605180.

\bibitem{Hosotani:2006nq}
Y. Hosotani,
\newblock (2006), hep-ph/0607064.

\bibitem{vonGersdorff:2005ce}
G. von Gersdorff and A. Hebecker,
\newblock Nucl. Phys. B720 (2005) 211, hep-th/0504002.

\bibitem{Knechtli:2005dw}
F. Knechtli, B. Bunk and N. Irges,
\newblock PoS LAT2005 (2005) 280, hep-lat/0509071.

\bibitem{Irges:2006zf}
N. Irges and F. Knechtli,
\newblock (2006), hep-lat/0604006.

\bibitem{Fu:1983ei}
Y.K. Fu and H.B. Nielsen,
\newblock Nucl. Phys. B236 (1984) 167.

\bibitem{Dimopoulos:2006qz}
P. Dimopoulos, K. Farakos and S. Vrentzos,
\newblock (2006), hep-lat/0607033.

\bibitem{Dvali:1996xe}
G.R. Dvali and M.A. Shifman,
\newblock Phys. Lett. B396 (1997) 64, hep-th/9612128.

\bibitem{Laine:2004ji}
M. Laine, H.B. Meyer, K. Rummukainen and M. Shaposhnikov,
\newblock JHEP 04 (2004) 027, hep-ph/0404058.

\bibitem{Chandrasekharan:1996ih}
S. Chandrasekharan and U.J. Wiese,
\newblock Nucl. Phys. B492 (1997) 455, hep-lat/9609042.

\bibitem{Brower:1997ha}
R. Brower, S. Chandrasekharan and U.J. Wiese,
\newblock Phys. Rev. D60 (1999) 094502, hep-th/9704106.

\bibitem{Schlittgen:2000xg}
B. Schlittgen and U.J. Wiese,
\newblock Phys. Rev. D63 (2001) 085007, hep-lat/0012014.

\bibitem{Kubo:2001zc}
M. Kubo, C.S. Lim and H. Yamashita,
\newblock Mod. Phys. Lett. A17 (2002) 2249, hep-ph/0111327.

\bibitem{Scrucca:2003ra}
C.A. Scrucca, M. Serone and L. Silvestrini,
\newblock Nucl. Phys. B669 (2003) 128, hep-ph/0304220.

\bibitem{Panico:2005dh}
G. Panico, M. Serone and A. Wulzer,
\newblock Nucl. Phys. B739 (2006) 186, hep-ph/0510373.

\bibitem{ZinnJustin-Book}
J. Zinn-Justin,
\newblock Quantum Field Theory and Critical Phenomena, 4th ed. (International
  Series of Monographs on Physics --- Vol. 113, Clarendon Press, Oxford, 2002).

\bibitem{Montvay:1984wy}
I. Montvay,
\newblock Phys. Lett. B150 (1985) 441.

\bibitem{Ejiri:2000fc}
S. Ejiri, J. Kubo and M. Murata,
\newblock Phys. Rev. D62 (2000) 105025, hep-ph/0006217.

\bibitem{Arnold:2002jk}
G. Arnold, B. Bunk, T. Lippert and K. Schilling,
\newblock Nucl. Phys. Proc. Suppl. 119 (2003) 864, hep-lat/0210010.

\bibitem{Fradkin:1978dv}
E.H. Fradkin and S.H. Shenker,
\newblock Phys. Rev. D19 (1979) 3682.

\bibitem{Osterwalder:1977pc}
K. Osterwalder and E. Seiler,
\newblock Ann. Phys. 110 (1978) 440.

\bibitem{Jansen:1985cq}
K. Jansen, J. Jersak, C.B. Lang, T. Neuhaus and G. Vones,
\newblock Phys. Lett. B155 (1985) 268.

\bibitem{Symanzik:1981hc}
K. Symanzik,
\newblock Mathematical Problems in Theoretical Physics, eds. R. Schrader et
  al., Lecture Notes in Physics 153 (1982) 47,
\newblock Presented at 6th Int. Conf. on Mathematical Physics, Berlin, West
  Germany, Aug 11-21, 1981.

\bibitem{Symanzik:1983dc}
K. Symanzik,
\newblock Nucl. Phys. B226 (1983) 187.

\bibitem{Symanzik:1983gh}
K. Symanzik,
\newblock Nucl. Phys. B226 (1983) 205.

\bibitem{Luscher:1998pe}
M. L{\"u}scher,
\newblock (1998), hep-lat/9802029.

\bibitem{Luscher:1990ck}
M. Luscher and U. Wolff,
\newblock Nucl. Phys. B339 (1990) 222.

\bibitem{Montvay:1987us}
I. Montvay and P. Weisz,
\newblock Nucl. Phys. B290 (1987) 327.

\bibitem{Knechtli:1999tw}
F. Knechtli,
\newblock (1999), hep-lat/9910044.

\bibitem{Albanese:1987ds}
APE, M. Albanese et~al.,
\newblock Phys. Lett. B192 (1987) 163.

\bibitem{Creutz:1979dw}
M. Creutz,
\newblock Phys. Rev. Lett. 43 (1979) 553.

\bibitem{Beard:1997ic}
B.B. Beard et~al.,
\newblock Nucl. Phys. Proc. Suppl. 63 (1998) 775, hep-lat/9709120.

\bibitem{McLerran:1980pk}
L.D. McLerran and B. Svetitsky,
\newblock Phys. Lett. B98 (1981) 195.

\bibitem{Kuti:1980gh}
J. Kuti, J. Polonyi and K. Szlachanyi,
\newblock Phys. Lett. B98 (1981) 199.

\bibitem{Dienes:1998vg}
K.R. Dienes, E. Dudas and T. Gherghetta,
\newblock Nucl. Phys. B537 (1999) 47, hep-ph/9806292.

\end{thebibliography}
